\documentclass[conference]{IEEEtran}

\usepackage{amsmath,amssymb,comment}

\usepackage{url}
\usepackage{graphicx}
\usepackage{subfigure}
\usepackage{color}
\usepackage{epsfig}
\usepackage{latexsym}

\usepackage{mathtools}

\usepackage{amsmath,amssymb,amsfonts}
\usepackage{mathtools}
\usepackage{dsfont} 
\usepackage{array}
\newcolumntype{L}{>{\centering\arraybackslash}m{1.7cm}}
\newcolumntype{M}{>{\centering\arraybackslash}m{1.7cm}}
\newcolumntype{N}{>{\centering\arraybackslash}m{3.808cm}}

\newtheorem{theorem}{Theorem}[section]
\newtheorem{definition}[theorem]{Definition}
\newtheorem{lemma}[theorem]{Lemma}
\newtheorem{proposition}[theorem]{Proposition}
\newtheorem{corollary}[theorem]{Corollary}

\newcommand{\fsquare}{\vrule height6pt width7pt depth1pt}   
\newcommand{\myproof}{{\hfill  \\ \bf Proof. \ }}           
\newcommand{\myendpf}{\hfill\fsquare \\[0.1in]}

\DeclarePairedDelimiter\floor{\lfloor}{\rfloor}

\usepackage{cite}

\allowdisplaybreaks
\def \OO {\mathrm{O}}
\def \oo {\mathrm{o}}
\newcommand{\pr}{\mathbb{P}}
\newcommand{\E}{\mathbb{E}}
\newcommand{\N}{\mathbb{N}}
\newcommand{\R}{\mathbb{R}}

\newcommand{\kk}{\langle K_{n} \rangle}
\newcommand{\K}{ K_{n}}

\newcommand{\limit}{\underset{n \rightarrow \infty}\lim}
\newcommand{\ii}{\mathds{1}} 
\newcommand{\comp}{^{\rm c}}

\newcommand{\hh}{\mathbb{H}(n;\mu,K_n)} 
\newcommand{\HH}{\mathbb{H}(n;\pmb{\mu},\pmb{K}_n)} 

\newcommand{\nodes}{\mathcal{N}}
\newcommand{\iii}{\mathbb{I}(n;\mu,K_n)} 
\newcommand{\hhc}{\mathbb{H}(n;\mu,\tilde{K}_n)}
\newcommand{\hhgc}{\mathbb{H}(n;\tilde{\mu},{K}_{r,n})}
\newcommand{\cmax}{C_{\rm max} (n;\mu, K_n)}
\newcommand{\kv}{\kappa_v}
\newcommand{\g}{\gamma}
\newcommand{\dut}{\mathcal{D}_{\rm UT}}
\newcommand{\ct}{\mathcal{C}_{\rm T}}
\newcommand{\but}{\mathcal{B}_{\rm UT}}

\newcommand{\aut}{\mathcal{A}_{\rm U,T}}
\newcommand{\alr}{\mathcal{A}_{\rm \ell,r}}
\newcommand{\blr}{\mathcal{B}_{\rm \ell,r}}
\newcommand{\clr}{\mathcal{C}_{\rm \ell,r}}
\newcommand{\dlr}{\mathcal{D}_{\rm \ell,r}}

\newcommand{\sdcon}{\mathcal{E}_{n,r}(\mu, K_n)}
\newcommand{\dst}{D_{\rm ST}}
\newcommand{\dts}{D_{\rm TS}}
\newcommand{\eij}{E_{ ij}}
\newcommand{\tone}{\mathcal{X}_1}
\newcommand{\ttwo}{\mathcal{X}_2}
\newcommand{\type}{\mathcal{X}_t}
\newcommand{\tu}{\vec{t}_{\rm u}}

\begin{document}

\title{On the Strength of Connectivity of Inhomogeneous Random K-out Graphs}

\author{\IEEEauthorblockN{Mansi Sood and Osman Ya\u{g}an}
\IEEEauthorblockA{Department
of Electrical and Computer Engineering and CyLab, \\
Carnegie Mellon University, Pittsburgh,
PA, 15213 USA\\
msood@andrew.cmu.edu, oyagan@ece.cmu.edu}}

\maketitle

\begin{abstract}
Random graphs are an important tool for modelling and analyzing the underlying properties of complex real-world networks. In this paper, we study a class of random graphs known as the {\em inhomogeneous random K-out graphs} which were recently introduced to analyze heterogeneous networks. In this model, first, each of the $n$ nodes is classified as type-1 (respectively, type-2) with probability $0<\mu<1$ (respectively, $1-\mu)$ independently from each other. Next, each type-1 (respectively, type-2) node draws 1  {\em arc} towards a node (respectively, $K_n$ arcs towards $K_n$ distinct nodes) selected uniformly at random, and then the orientation of the arcs is ignored. A main design question is how should the parameters $n$, $\mu$, and $K_n$ be selected such that the network exhibits certain desirable properties with high probability. Of particular interest is the {\em strength} of connectivity often studied in terms of $k$-connectivity; i.e., with $k=1, 2, \ldots$, the property that the network remains connected despite the removal of any $k-1$ nodes or links. When the network is {\em not} connected, it is of interest to analyze the {\em size} of its {\em largest} connected sub-network. In this paper, we answer these questions by analyzing the {\em inhomogeneous random K-out graph}. From the literature on {\em homogeneous} K-out graphs wherein all nodes select $K_n$ neighbors (i.e., $\mu=0$), it is known that when $K_n \geq2$, the graph is $K_n$-connected asymptotically almost surely (a.a.s.) as $n$ gets large. In the inhomogeneous case (i.e., $\mu>0$), it was recently established that achieving even 1-connectivity a.a.s. requires $K_n=\omega(1)$. Here, we provide a comprehensive set of results to complement these existing results. First, we establish a sharp {\em zero-one law} for $k$-connectivity, showing that for the network to be  $k$-connected a.a.s., we need to set 
$K_n = \frac{1}{1-\mu}(\log n +(k-2)\log\log n + \omega(1))$
for all $k=2, 3, \ldots$. Despite such large scaling of $K_n$ being required for $k$-connectivity,  we show that the trivial condition of $K_n \geq 2$ for all $n$ is sufficient to ensure that inhomogeneous K-out graph has a connected component of size $n-\OO(1)$ whp. Put differently, even with $K_n =2$, all but finitely many nodes will form a connected sub-network under any $\mu,\ 0<\mu<1$. We  present an upper  bound on the probability that more than $M$ nodes are outside of the largest component, and show that this decays  as $O(1)\exp\{-M(1-\mu)(K_n-1)\} +o(1)$. Through numerical experiments, we demonstrate the usefulness of our results when the number of nodes is finite.


\end{abstract}

{\bf Keywords:} Random Graphs, Inhomogeneous Random K-out Graphs, Giant component, Connectivity, Security.

\section{Introduction}
\label{sec:introduction}

Random graph modeling is an important framework for developing fundamental insights into the structure and dynamics of several complex real-world networks including social networks, economic networks and communication networks\cite{boccaletti2006complex, goldenberg2010survey, newman2002random, kakade2005economic}. In the context of wireless sensor networks (WSNs), random graph  models have been used widely \cite{Gligor_2002,yagan2011random} in the design and performance evaluation of {\em random key predistribution schemes}, which were proposed for ensuring secure connectivity \cite{Gligor_2002,security_survey,XiaoSurvey}. 
In recent years, the analysis of heterogeneous variants of classical random graph models has emerged as an important topic 
\cite{eletrebycdc2018,eletreby2019ISIT,du2007effective,8606999, Rashad/Inhomo, Yagan/Inhomogeneous}, 
owing to the fact that real-life network applications are increasingly {\em heterogeneous} with participating nodes having different capabilities and (security and connectivity) requirements \cite{Barabasi_1999, boccaletti2006complex, Lu2008_applications, Wu2007_applications, Yarvis_2005}.

Random K-out graph  is one of the earliest models studied in the literature \cite{FennerFrieze1982,Bollobas}. 
Denoted here by $\mathbb{H}(n;K)$ it is constructed as follows. Each of the $n$ nodes draws $K$ arcs towards $K$ distinct nodes chosen uniformly at random among all others. 
The orientation of the arcs is then ignored,
yielding an {\em undirected} graph.
Recently, random K-out graphs have been studied \cite{Yagan2013Pairwise,yagan2012modeling,yavuz2017k,yavuz2015toward}
in the context of the random {\em pairwise} key predistribution scheme \cite{Haowen_2003}; along with the original key predistribution scheme proposed by Escheanuer and Gligor \cite{Gligor_2002}, the pairwise scheme is one of the most widely recognized security protocols for WSNs. 
Another recent application of random K-out graphs is the {\em Dandelion} protocol proposed by Fanti et al.  \cite[Algorithm~1]{FantiDandelion2018},
where a similar structure was used for   message diffusion that is robust to {\em de-anonymization} attacks. Of particular interest to this work is the connectivity of random K-out graphs. It was established in  \cite{Yagan2013Pairwise, FennerFrieze1982} that random K-out graphs are connected (respectively, not connected) {\em with high probability} (whp) when $K \geq 2$ (respectively, when $K=1$); i.e., 
\begin{equation} 
\lim_{n \to \infty} \mathbb{P}\left[ \mathbb{H}(n;K) \text{ is connected}\right] =
\begin{cases}
1 & \mathrm{if} \quad K\geq 2, \\
0 & \mathrm{if} \quad K=1.
\end{cases}
\label{eq:homogeneous_zero_one_law}
\end{equation}

\begin{figure}
\centering
\includegraphics[scale=0.11]{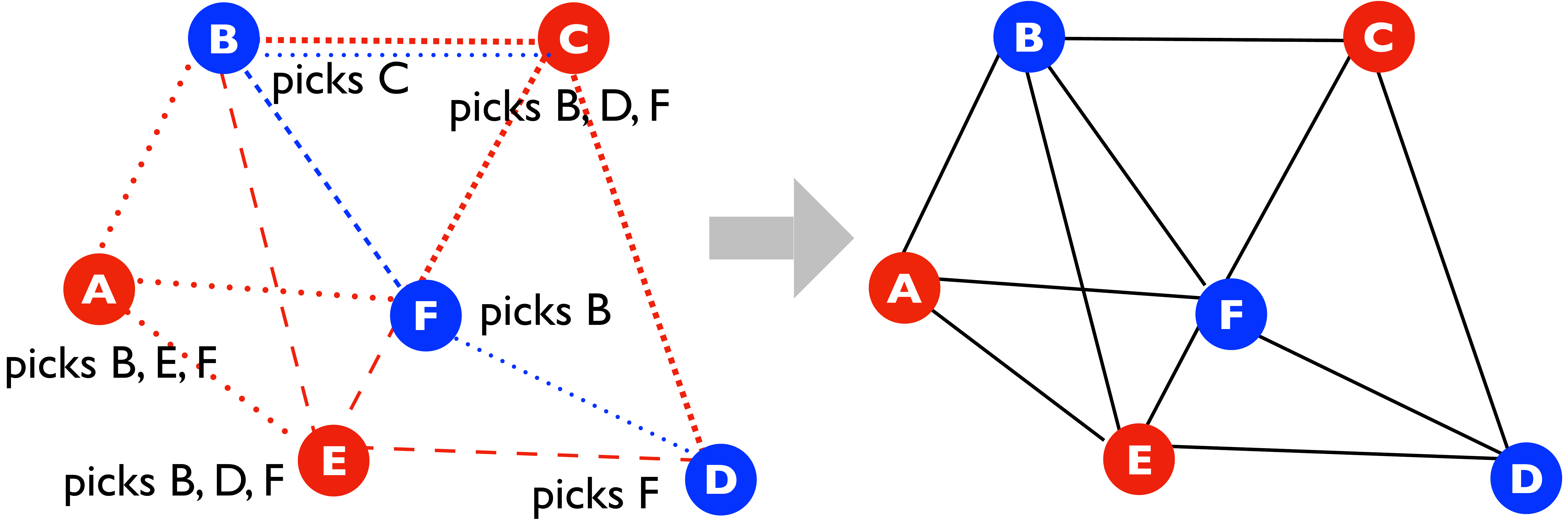} 
\caption{
An inhomogeneous random K-out graph with $6$ nodes. 
Nodes $A, C$ and $E$ are type-2 and the rest ($B,D,F$) are type-1.
Each type-1 (resp. type-2) node selects 1 (resp. $K_n=3$) node uniformly at random.  An edge is drawn between two nodes if at least one selects the other.}
\vspace{-2mm}
\label{fig:pairwise}
\end{figure}

Motivated by the aforementioned emergence of heterogeneity in many real-life networks, Eletreby and Ya\u{g}an studied \cite{eletrebycdc2018} the {\em inhomogeneous} random K-out graph. Therein, each node is classified as type-1 (respectively, type-2) with probability $\mu$ (respectively, $1-\mu$).
Then, each type-1 (respectively, type-2) node selects one node (respectively, $K_n \geq 2$ nodes) uniformly at random from all other nodes; 
see Figure~\ref{fig:pairwise}.  Here, the notation $K_n$ indicates that the number of selections made by type-2 nodes {\em scales} as a function of the number of nodes $n$. 

 In \cite{eletrebycdc2018},
 it was shown that for any $0<\mu<1$, 
 the inhomogeneous random K-out graph is connected whp if and only if $K_n$ grows unboundedly large with $n$; i.e.,
\begin{equation} 
\lim_{n \to \infty} \hspace{-.5mm}\mathbb{P}\left[ \mathbb{H}(n;\mu,K_n) \text{ is connected}\right] \hspace{-.5mm}=\hspace{-.5mm}
\begin{cases}
1 & \mathrm{\hspace{-1.5mm}if}~  {\color{black}K_n \rightarrow \infty} \\
<1 & \mathrm{\hspace{-1.5mm}otherwise}. 
\end{cases}
\label{eq:introEqcdc}
\end{equation}  

This paper complements (\ref{eq:introEqcdc}) through a comprehensive set of results concerning the {\em strength} of connectivity in inhomogeneous K-out graphs. \emph{First}, we focus on $k$-connectivity of  the inhomogeneous random K-out graph. The notion of $k$-connectivity used in this paper coincides with $k$-vertex connectivity, which is defined as the property that the graph remains connected after deletion of any $k-1$ vertices.  It is known that a $k$-vertex connected graph is always $k$-edge connected, meaning that it will remain connected despite the removal of any $k-1$ edges \cite{erdos61conn},\cite[p.~11]{diestel2006graph}. Thus, we say that a graph is $k$-connected (without explicitly referring to vertex-connectivity) to refer to the fact that it will remain connected despite the deletion of any $k-1$ vertices or edges. By Menger's Theorem \cite[p.~50~Theorem 3.3.1]{diestel2006graph}, it is known that if a network is $k$-connected, then there exist at least $k$ disjoint paths between  all pairs of nodes.

\begin{figure}
\hspace{-.3cm}
\centering
\includegraphics[scale=0.12]{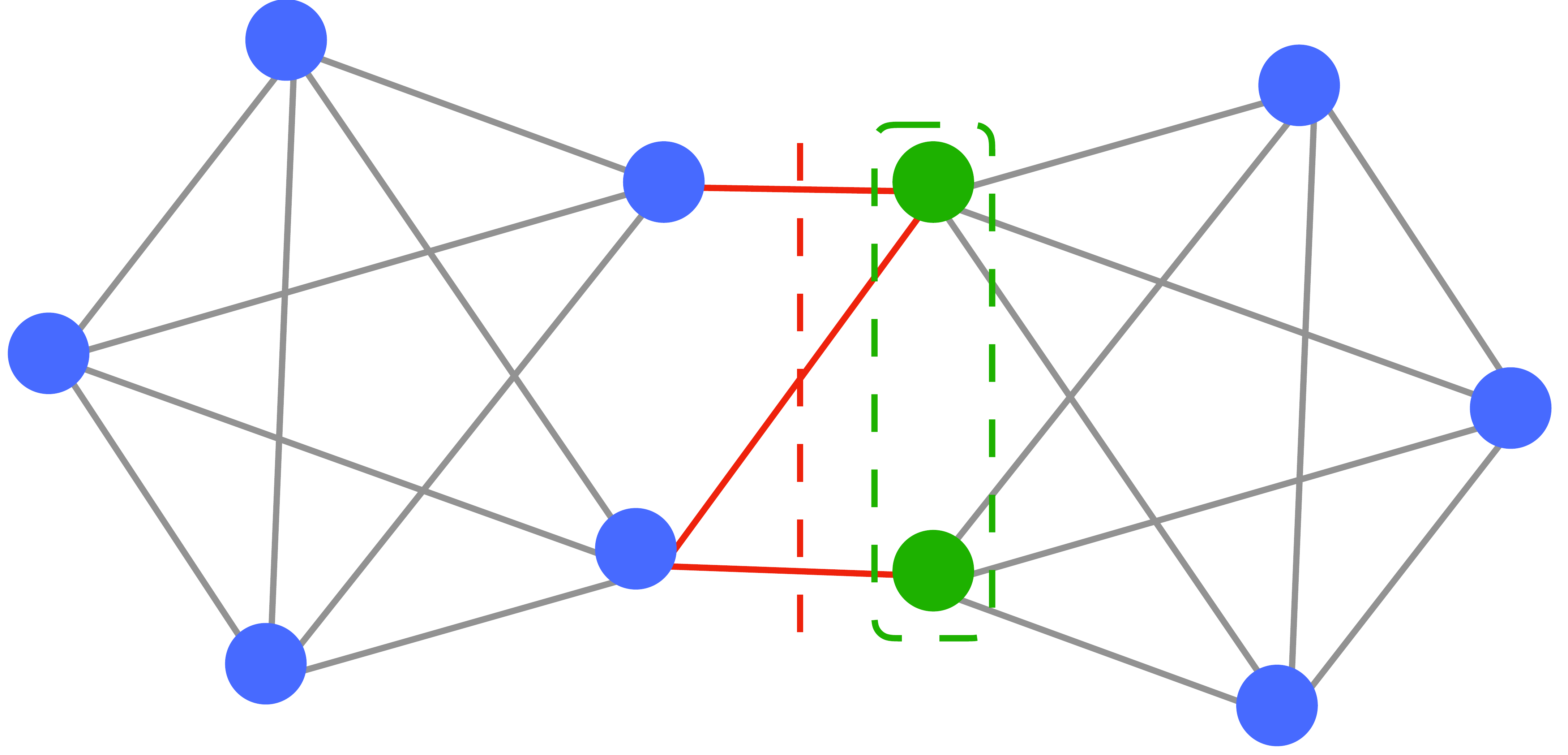} 
\caption{Graph $\mathcal{G}$ has 10 nodes and a minimum node degree of $4$. There exists an edge cut of size $3$ and a vertex cut of size $2$ indicated in red and green, respectively, which can be deleted to disconnect $\mathcal{G}$. Here, $\mathcal{G}$ is $3$-edge connected and $2$-vertex connected. This illustrates that a graph with minimum node degree $k$ is neither $k$-edge nor $k$-vertex connected in general; see \cite[p.~43]{bondy1976graph} for more details.}
\vspace{-2mm}
\label{fig:kconmndnotsame}
\end{figure}

In the context of WSNs, the property of $k$-connectivity is highly desirable since it provides higher degree of fault tolerance and information accuracy in aggregating information from multiple sensors \cite{liu2006coverage}. As a first step towards proving $k$-connectivity, the authors analyzed  \cite{extvs} the minimum node degree of $\hh$ and established that for any $k\geq2$,
\begin{align}
  \limit \pr\left[\begin{array}{ll}
 \text{Min.~node degree of }    &  \\ 
     \text{$\hh$ is  $\geq k$} & 
 \end{array}
 \hspace{-4mm}\right]
  = \begin{cases}
                                 1 & \textrm{if } \limit \g_n =+\infty, \\ 
                                  0 & \textrm{if } \limit \g_n =-\infty, \\
 \end{cases}
  \nonumber
\end{align}
where $\g_n$ is defined through $\K=\left(\frac{\log n +(k-2)\log \log n }{1-\mu}\right)+\g_n$.
The property of $k$-connectivity requires a minimum node degree of at least $k$ as a necessary condition. However, it is important to note that \emph{a minimum node degree being at least $k$ does not guarantee $k$-connectivity}. 
The minimum node degree being at least $k$ does not even ensure a weaker property of $k$-edge connectivity.

This is illustrated in {Figure~\ref{fig:kconmndnotsame}} where the example graph  $\mathcal{G}$ has a minimum degree of $4$,  but can be made disconnected by removing 2 vertices or 3 edges. In this example, $\mathcal{G}$ is only 2-vertex-connected and 3-edge connected \cite[p.~11]{diestel2006graph}.  
In \cite[Conjecture 2]{extvs}, we conjectured that 
taking evidence from several other random graph models \cite{erdos61conn,ZhaoYaganGligor, PenroseBook}, there would exist a zero-one law for $k$-connectivity analogous to the zero-one law for the minimum node degree being at least $k$. In this work, we prove
that this conjecture indeed holds.
We derive scaling conditions on $\mu,\K$ such that the  inhomogeneous random K-out graph is $k$-connected asymptotically almost surely as $n$ gets large, where $k=2,3,\dots$. We present our result in terms of a sharp zero-one law. 
For any $k \geq 2$, we show that if $K_n = \frac{1}{1-\mu}(\log n +(k-2)\log\log n + \omega(1))$,
then  $\hh$ is $k$-connected asymptotically almost surely (a.a.s.). In contrast, if $K_n = \frac{1}{1-\mu}(\log n +(k-2)\log\log n - \omega(1))$, then $\hh$ is a.a.s. {\em not} $k$-connected. This result shows that if there is a positive fraction of type-1 nodes, then  type-2 nodes must make  $K_n=\Omega(\log n)$ selections for the network to achieve $k$-connectivity for any $k =2, 3, \ldots$. This is rather unexpected given that the network is a.a.s. 1-connected under any $K_n= \omega(1)$. The result  is also in contrast with most other random graph models where the zero-one law for $k$-connectivity appears in a form that reduces to a zero-one law for 1-connectivity by simply setting $k=1$.
Through simulations we study the impact of the parameters ($\mu, \K$) on the probability of $k$-connectivity when the number of nodes is finite and observe an agreement with our asymptotic results.

The heterogeneity of node types makes $\hh$ a complicated model and the proofs involve techniques that are different from those used for the homogeneous K-out random graph \cite{FennerFrieze1982}, \cite{Yagan2013Pairwise}. Moreover, the proof for this case varies significantly from results on $1$-connectivity for inhomogeneous random K-out graphs \cite{eletrebycdc2018} and uses new tools including {\em conditional} negative association (of certain random variables of interest) introduced recently in \cite{yuan2010conditional}.

As seen from (\ref{eq:introEqcdc}), ensuring connectivity of  $\hh$ requires $\K=\omega(1)$. Although it is desirable to have a  connected network, in several practical applications, resource constraints can potentially limit the number of links that can be successfully established \cite{hwang2004revisiting}.
In such scenarios, it may suffice to have a large connected sub-network spanning almost the entire network \cite{MeiPanconesiRadhakrishnan2008}
depending on the application.
 For example, if a sensor network is designed to monitor temperature of a field, then instead of knowing the temperature at every location in the field, it may suffice to have readings from a majority of sensors in the field \cite{liu2006coverage}.

With this in mind, the \emph{second} question which we address here is when $\K$ is bounded (i.e., $K_n=O(1)$), how many nodes are contained in the largest connected sub-network (i.e., component) of $\hh$?
In the literature of random graphs, this is often studied in terms of the {\em emergence} and size of the {\em giant component}, defined as a connected sub-network comprising $\Omega(n)$ nodes; see \cite{erdHos1960evolution} for a classical example on the giant component of Erd\H{o}s-R\'enyi graphs. 

Here, we show that the inhomogeneous random K-out graph contains a giant component as long as  the trivial conditions $0<\mu<1$ and $\K \geq 2$ (for all $n$) hold. In fact, we show that under the same conditions, the graph contains a connected sub-network of size $n- \OO(1)$ whp. Put differently, all but finitely many nodes will be contained in the giant component of $\hh$, as $n$ goes to infinity whp. This is also demonstrated through numerical experiments where we observe that   with $n=5000, \mu=0.9, \K=2$,  at most 45 nodes turned out to be  outside the largest connected component across 100,000  experiments; see Section~\ref{sec:Main Results} for details.

Our result on the giant component follows from an upper bound on the probability that more than $M$ nodes are outside of the giant component. We show  that this probability decays at least as fast as  $O(1)\exp\{-M(1-\mu)(K_n-1)\} +o(1)$ providing a clear trade-off between $K_n$ and the fraction $(1-\mu)$ of nodes that make $K_n$ selections. Our proof technique deviates from most of the works on the size of the giant component that are based on analyzing a branching process. Instead, we rely on a simpler approach based on the connection between the {\em non-existence} of sub-graphs with size exceeding $M$ and that are {\em isolated} from the rest of the graph, and the size of the of largest component being at least $n-M$.

We close by describing a potential future application of (inhomogeneous) random K-out graphs. Given their sparse yet connected structure, this model can be useful for analyzing payment channel networks (PCNs) wherein edges represent the funds escrowed in a bidirectional overlay network on top of the cryptocurrency network \cite{lightning2016}. Recent work in the realm of cryptocurrency networks has closely looked at the topological properties of PCNs and their impact on the achieved throughput \cite{pcntopology2019,tang2019privacy, sivaraman2018high}. A key aspect of PCNs is the trade-off between the number of edges in the network (which is constrained since  funds need to be committed on each edge) and its  connectivity (which is desirable so that any pair of nodes can perform transactions with each other). The results established here show that the construction of inhomogeneous random K-out graphs leads to almost all nodes being connected with each other (as part of the largest connected component) with relatively small number of edges per node; e.g., with $K=2$ and $\mu=0.5$, each node will have 3 edges on average. In fact,  the
Lightning Network dataset from December 2018 shows that it contains 2273 nodes, of which 2266  are contained in the largest connected component while the remaining 7 nodes being in three isolated components.

All limits are understood with the number of nodes $n$ going to infinity. While comparing asymptotic behavior of a pair of sequences $\{a_n\},\{b_n\}$, we use $a_n = \oo(b_n)$, $a_n=\omega(b_n)$,  $a_n = \OO(b_n)$, $a_n=\Theta(b_n)$, {\color{black}and $a_n = \Omega(b_n)$} with their meaning in the standard Landau notation. 
All random variables are defined on the same probability triple $(\Omega, {\mathcal{F}}, \mathbb{P})$.
Probabilistic statements are made with respect to this probability measure $\mathbb{P}$, and we denote the corresponding expectation operator by $\mathbb{E}$. For an event $A$, its complement is denoted by $A\comp$. We let $\ii[A]$ denote the indicator random variable which takes the value 1 if event $A$ occurs and 0 otherwise. 
 We say that an event occurs with high probability (whp) 
 if it holds with \emph{probability tending to one} as $n\rightarrow \infty$. We denote the cardinality of a discrete set $A$ by $|A|$ and the set of all positive integers by $\N_0$. For events $A$ and $B$,  we use $A \implies B$ with the meaning that  $A \subseteq B$.
 
\section{Inhomogeneous Random K-out graph}
\label{sec:model}

Let $\nodes:=\{1,2,\dots,n\}$ denote the set of vertex labels 
and let $\nodes_{-i}:=\{1,2, \dots, n\}\setminus i$. 
In its simplest form, the inhomogeneous random K-out graph is constructed on the vertex set $\{v_1, \ldots, v_n\}$  as follows. First, each vertex is assigned as type-1 (respectively, type-2) with probability $\mu$ (respectively, $1-\mu$) independently from other nodes, where $0<\mu<1$. 
Next, each type-1 (respectively, type-2) node selects $K_1$ (respectively, $K_2$) distinct nodes  uniformly at random among all other nodes. For each $i \in \nodes$, let $\Gamma_{n,i} \subseteq \nodes_{-i}$ denote the labels corresponding to the selections made by $v_i$.  Under the aforementioned assumptions,  $\Gamma_{n,1}, \ldots , \Gamma_{n,n}$ are mutually independent {\em given} the types of nodes. 
We say that  distinct nodes $v_i$ and $v_j$ are adjacent, denoted by $v_i \sim v_j$ if at least one of them picks the other. Namely, 
\begin{align}
v_i \sim v_j ~~\quad \mbox{if} ~~~\quad j \in \Gamma_{n,i} ~\vee~ i \in \Gamma_{n,j}. 
\label{eq:Adjacency}
\end{align}
The inhomogeneous random K-out graph is then defined on the vertices $\{v_1,\ldots,v_n\}$ through the adjacency condition (\ref{eq:Adjacency}). More general constructions with arbitrary number of node types is also possible \cite{eletreby2019ISIT}, and the implications of our results for such cases will be discussed later.

As in  \cite{eletrebycdc2018}, we assume that $K_1=1$ which in turn implies that $K_2 \geq 2$. We allow $K_2$ to scale with (i.e., to be a function of) $n$  and simplify the notation by denoting the corresponding mapping as $K_n$.  Put differently, we consider the inhomogeneous random K-out graph, denoted as $\hh$, where each of the $n$ nodes selects one other node with probability $0<\mu<1$ and $K_n$ other nodes with probability $1-\mu$; the edges are then constructed according to (\ref{eq:Adjacency}). Throughout, it is assumed that $K_n \geq 2$ for all $n$ in line with the assumption that $K_2 > K_1 =1$. We denote the average number of selections made by each node in $\hh$ by $\kk$. It is straightforward to see that
\begin{align}
 \kk =   \mu+(1-\mu)K_n.
 \label{eq:avg_K}
\end{align}

 



\section{Main Results: $k$-connectivity}
\label{sec:Main Results}
We refer to any mapping $K:\N_0 \rightarrow \N_0$ satisfying the conditions 
$2 \leq K_n < n$
for all $n=2, 3, \ldots$
as a \emph{scaling}. We say that a graph is $k$-connected  if it  remains connected despite the deletion of any $k-1$ vertices or edges. 
 Next, we present our first main result that characterizes the critical scaling of the parameters $(\mu,\K)$ under which the inhomogeneous random K-out graph $\hh$ 
 is $k$-connected asymptotically almost surely.

\begin{theorem}
{\sl 
Consider a scaling $K:\N_0 \rightarrow \N_0$ and $\mu$ such that $0<\mu<1$. 
With $\kk = \mu+(1-\mu)K_n$ and an integer $k\geq2$ let the sequence $\g: \N_0 \rightarrow \R$ be  defined through
\begin{align}
\kk=\log n +(k-2)\log \log n+\g_n, 
\label{eq:scaling_avg}
\end{align}
for all $n=2, 3, \ldots$. Then, we have
\begin{align}
   \limit \pr\left[\begin{array}{ll}
   \textrm{$\hh$ is }    &  \\
     \textrm{$k$-connected} & 
  \end{array}
  \hspace{-4mm}\right]
  = \begin{cases}
                                   1 & \textrm{if } \limit \g_n =+\infty, \\
                                   0 & \textrm{if } \limit \g_n =-\infty. \\
  \end{cases}
  \label{eq:zero_one_main_thm}
\end{align}
\label{theorem:kcon}
}
\end{theorem}
An outline of the proof of Theorem~\ref{theorem:kcon} is given in Section \ref{sec:kcon_outline}.  More details are presented in the Appendix. 

We note that  (\ref{eq:scaling_avg})  presents solely a definition of the sequence $\g_n$ without any loss of generality; it does not impose any assumption on the parameters ($\mu$, $\K$). 
The scaling condition (\ref{eq:scaling_avg}) could also be  expressed more explicitly in terms of $K_n$ as
\begin{align}
K_n=\frac{\log n +(k-2)\log \log n}{1-\mu}+\g_n \label{eq:kcon_scaling}
\end{align}
with the corresponding zero-one law 
(\ref{eq:zero_one_main_thm}) unchanged.

Theorem~\ref{theorem:kcon} provides a sharp zero-one law for the $k$-connectivity of the random graph $\hh$ as the size of the network grows large. In the context of WSNs, it establishes {\em critical} scaling conditions on the parameters of the pairwise scheme $(\mu,\K)$ under which the network will be securely and reliably connected whp. 
We see from \cite[Theorem~1]{extvs} that the critical scaling conditions for $k$-connectivity coincide with those for the minimum node degree to be at least $k$. 
This is similar to the case with most random graph models including Erd\H{o}s-R\'enyi (ER) graphs \cite{erdos61conn}, random key graphs \cite{ZhaoYaganGligor} and random geometric graphs \cite{PenroseBook}.

It follows from Theorem~\ref{theorem:kcon} that if there is a positive fraction $\mu$ of type-1 nodes, then  type-2 nodes must make  $K_n=\Omega(\log n)$ selections for the network to achieve $k$-connectivity for any $k =2, 3, \ldots$. As discussed below, this result is rather unexpected given that the network is a.a.s. 1-connected under any $K_n= \omega(1)$ as shown in \cite{Rashad/Inhomo}. This gap between 1-connectivity and $k$-connectivity for $k \geq 2$ is in contrast with most other random graph models where the zero-one law for $k$-connectivity appears in a form that reduces to a zero-one law for 1-connectivity by simply setting $k=1$; see more in Section \ref{subsec:discussion}.

\subsection{Discussion}
\label{subsec:discussion}

We discuss some implications of Theorems \ref{theorem:kcon} on the reliable connectivity of networks modelled by inhomogeneous random K-out graphs. 
With $\eij$ denoting the event that there exists an edge in $\hh$ between nodes $v_i$ and $v_j$, we have
\begin{align}
\pr[\eij]&= 1 - (1-\pr[i \in \Gamma_{n,j}])(1-\pr[j \in \Gamma_{n,i}]),\nonumber \\
&=1-\left(1-\dfrac{\kk}{n-1}\right)^2=\dfrac{2\kk}{n-1}-\left(\dfrac{\kk}{n-1}\right)^2.
\label{eq:kcon_meandegree}\end{align}
Thus, if $\kk=\oo( n)$, then the mean  degree in $\hh$ is $2\kk (1+o(1))$, while the mean  degree of type-1 nodes is $1+\kk$. Table~\ref{table:tab1} presents a comparison of the mean node degree needed for having $1$-connectivity and $k$-connectivity a.a.s. for homogeneous and inhomogeneous  random K-out graphs \cite{eletrebycdc2018, FennerFrieze1982}, and {\em random key graphs} \cite{yagan2012zero,Yagan/Inhomogeneous,8606999}. For  inhomogeneous models, the table entries correspond to the mean degree of the {\em least connected} node type. We also include the corresponding results for ER graphs \cite{erdos61conn} for comparison. 

 \begin{table}[!t]
\vspace{2mm}
\begin{center}
 \begin{tabular}{|L| M| N|} 
 \hline
\textbf{Random graph} & \textbf{$\pmb{1}$-connectivity} &   \textbf{$\pmb{k}$-connectivity}, $k\geq2$ \\
 \hline\hline
  Homogeneous K-out &  $4$ & $2k$ \\ 
 \hline
  Inhomogeneous K-out &  $\omega(1)$ & $\log n + (k-2) \log \log n + \omega(1)$  \\ 
 \hline
  Homogeneous random key &  $\log n + \omega(1)$ & $\log n + (k-1) \log \log n + \omega(1)$ \\
 \hline
  Inhomogeneous random key &  $\log n + \omega(1)$ & $\log n + (k-1) \log \log n + \omega(1)$ \\
 \hline
  Erd\H{o}s-R\'enyi &  $\log n + \omega(1)$ & $\log n + (k-1) \log \log n + \omega(1)$  \\
 \hline
\end{tabular}
\end{center}
\caption{\sl Mean node degree necessary for 1-connectivity and $k$-connectivity in several random graph models. For inhomogeneous K-out and inhomogeneous random key graphs, the values given in the table correspond to the mean degree for the {\em least} connected node type.}
\label{table:tab1}
\vspace{-7mm}
\end{table}

An interesting observation is that for the inhomogeneous random K-out graph, increasing the strength of connectivity from 1 to $k\geq2$  requires an increase of $\log n +(k-2) \log \log n$ in the mean degree. This is much larger than what is required (i.e., $(k-1) \log \log n$) in the other models seen in Table~\ref{table:tab1}. In fact, for most  random graph models, the zero-one law   for $1$ connectivity can be obtained from the corresponding result for $k$-connectivity by setting $k=1$; this can be confirmed from the entries in Table~\ref{table:tab1} for homogeneous/inhomogeneous random key graphs and ER graphs. To the best of our knowledge, inhomogeneous K-out graphs is the only model where the critical scalings for  1-connectivity and 2-connectivity differ significantly.

\begin{figure*}[!t]
\centering
\includegraphics[scale=0.22]{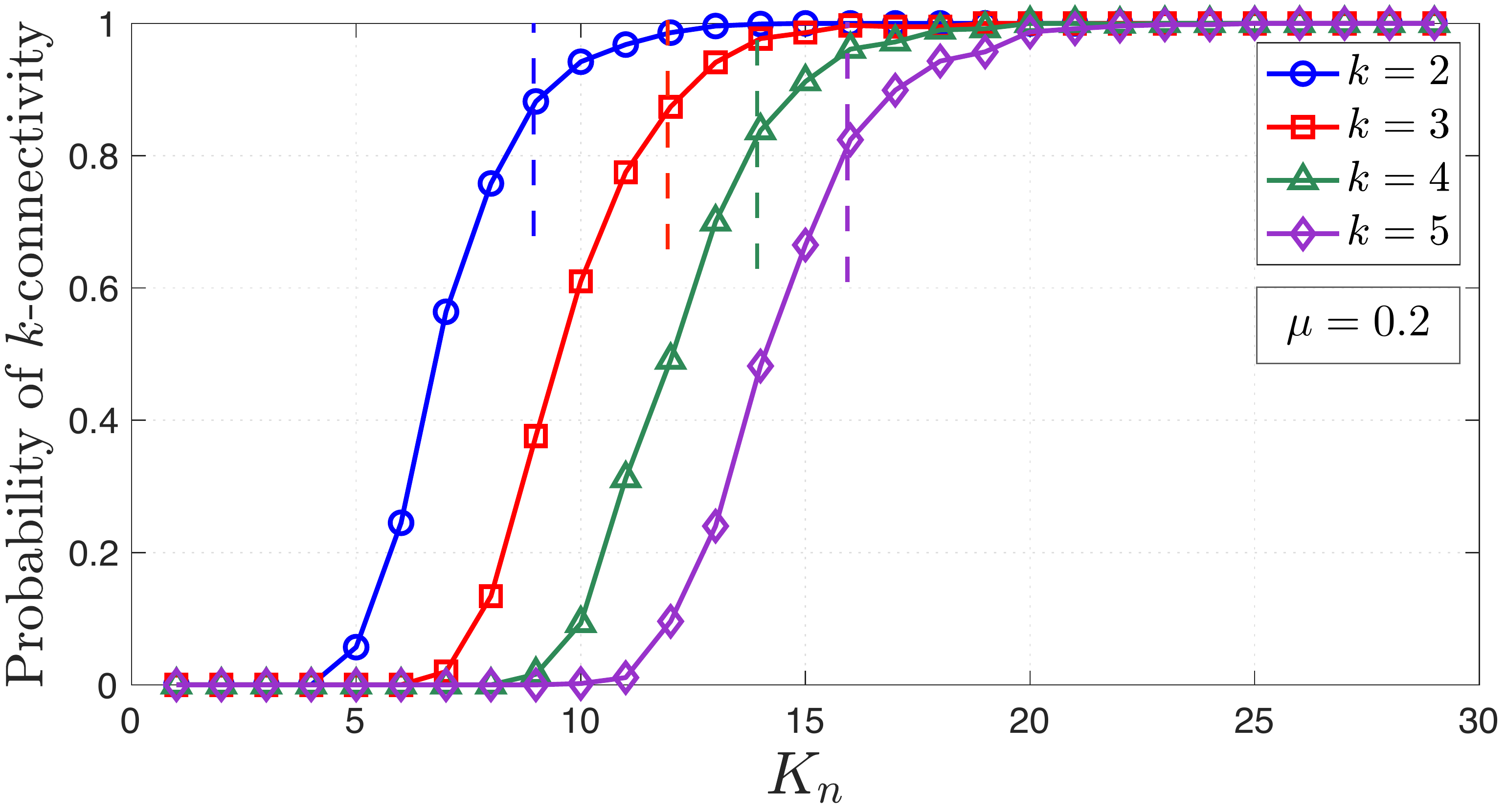}\label{fig:kconmu2}
\hspace{2cm}
\includegraphics[scale=0.22]{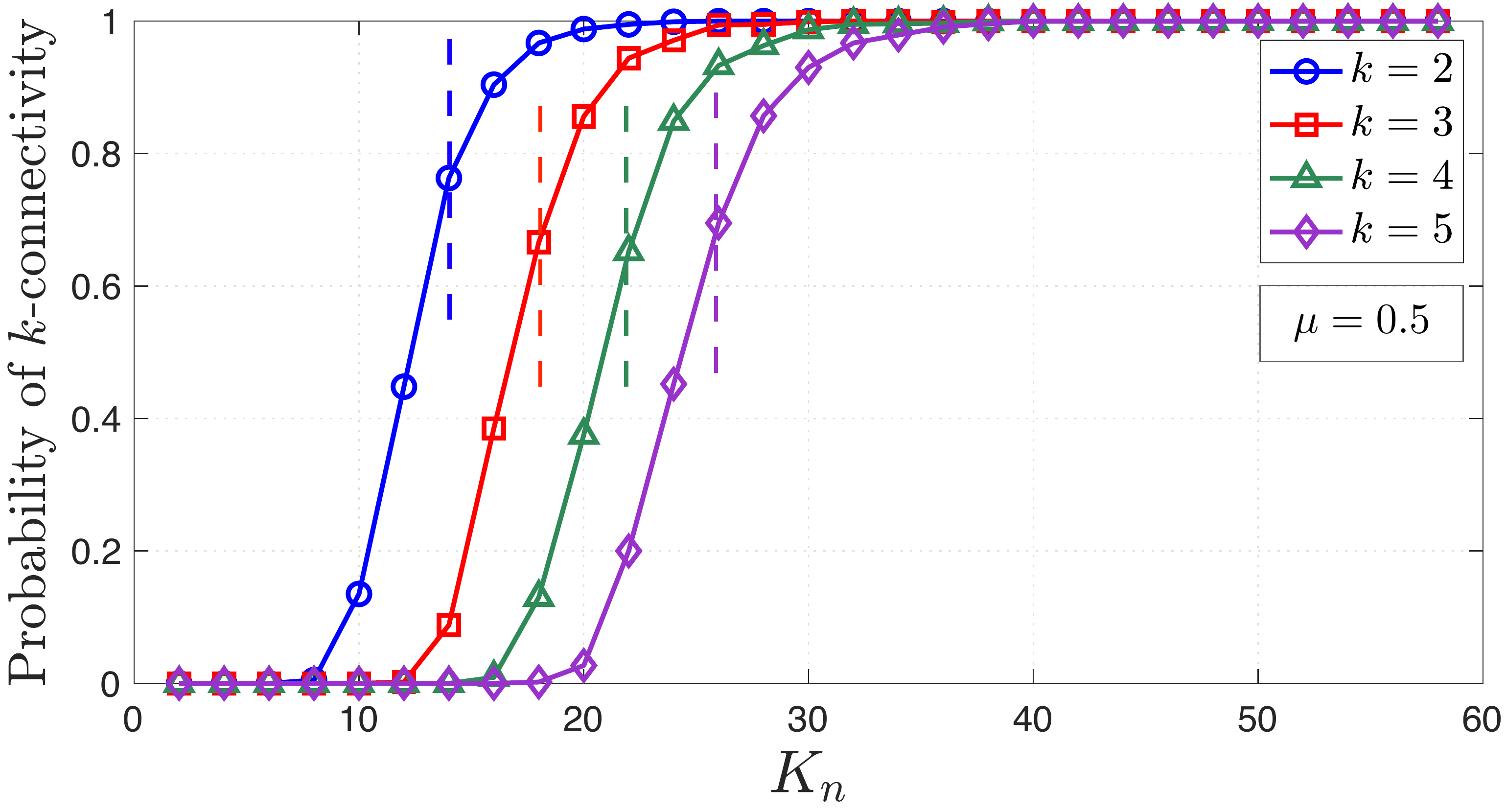}\label{fig:kconmu5}
\\ \vspace{3mm}
\includegraphics[scale=0.22]{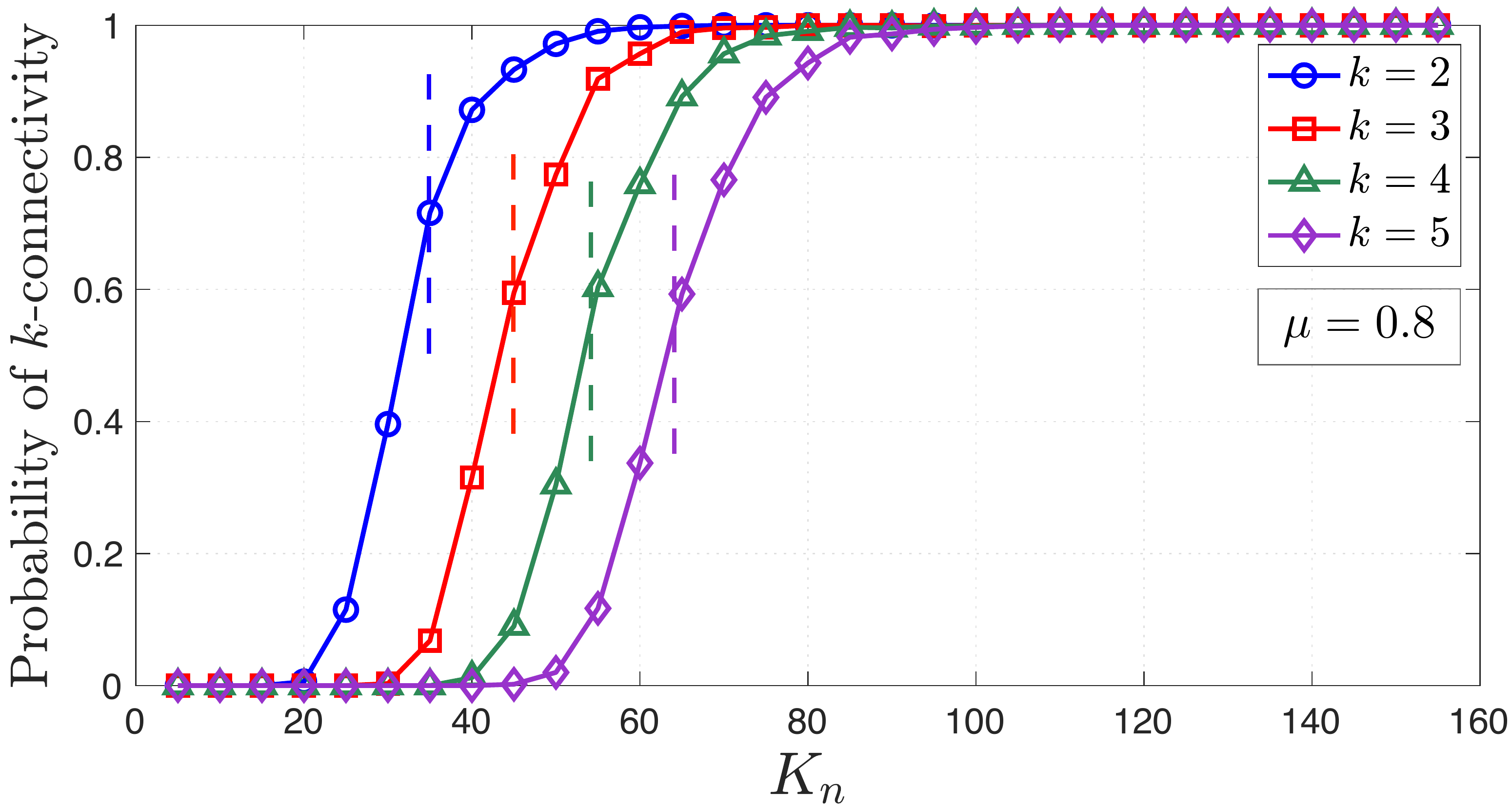}\label{fig:kconmu8}    
\hspace{2cm}
\includegraphics[scale=0.22]{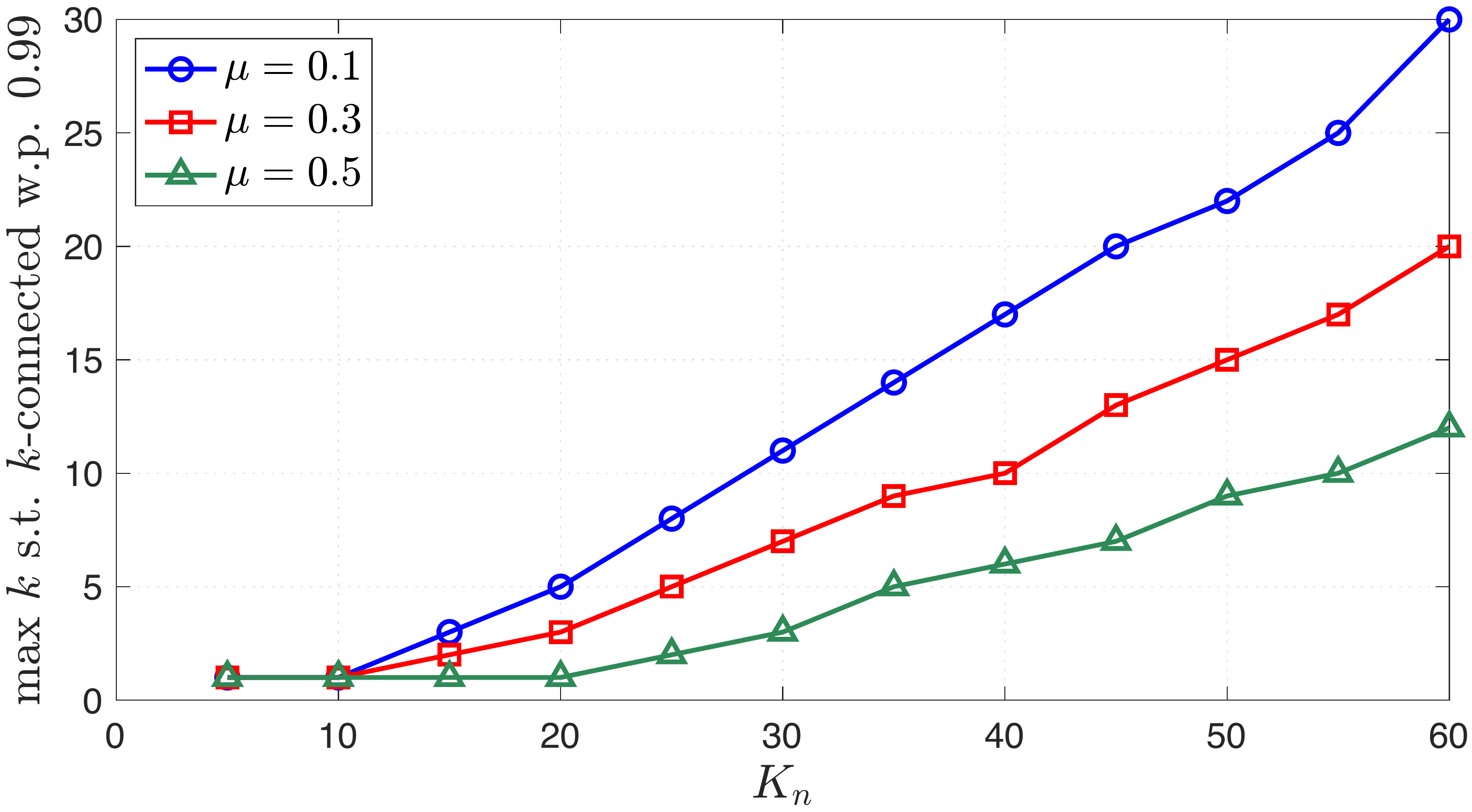}\label{fig:kvsK}   
\vspace{-2mm}
\caption{Empirical probability of $k$-connectivity of $\hh$ averaged over $1,000$ experiments  when $n=1000$, varying $K_n$ and three different $\mu$ values; $K_n$ is the number of choices made by type-2 nodes and $\mu$ is the fraction of type-1 nodes. The vertical dashed lines in indicate the threshold $K_n= \left \lceil \frac{\log n +(k-2)\log \log n}{1-\mu}
\right \rceil$ corresponding to the scaling condition (\ref{eq:kcon_scaling}) in Theorem~\ref{theorem:kcon}. The last plot shows the maximum value of $k$ such that $\hh$ is $k$-connected with probability (w.p.) at least 0.99.}
\vspace{-4mm}
 \label{fig:kcon}
\end{figure*}


\subsection{Numerical Results}
\label{sec:kcon_simulations}
We present simulation results to show the impact of the number of choices made by type-2 nodes ($\K$) and the probability of a node being assigned type-1 ($\mu$) on the probability that the resulting WSN is $k$-connected.
We consider an inhomogeneous random K-out graph comprising of $n=1000$ nodes. We first fix parameters $\mu =0.2,0.5,0.8$ and vary $K_n$. For each parameter tuple $(n,\mu,\K, k)$, 1000 independent realizations for $\hh$ are generated and empirical probability of $k$-connectivity is plotted in Figure~\ref{fig:kcon}.

A smaller value of $\mu$ corresponds to a network dominated by type-2 nodes. Consequently, for a low $\mu$ regime, the resulting graph is more dense and we expect to see stronger connectivity. Conversely, when $\mu$ is large, it takes a higher value for the parameter $\K$ to achieve the same strength of connectivity. This trend is reflected in Figure~\ref{fig:kcon} wherein the minimum $K_n$ required to make the network $k$-connected whp increases as $\mu$ increases. We point out that the scale of the plots for different $\mu$ has been chosen differently for compactly reporting roughly the same number of values of $K_n$ on either side of the phase transition. Whenever a network is $k$-connected, it automatically implies that the network is $\ell$-connected for all $\ell < k$. This manifests as the upward shift in the probability of $k$-connectivity as $k$ decreases in Figure~\ref{fig:kcon}.

The vertical dashed lines seen in Figure~\ref{fig:kcon} correspond to the {\em critical} thresholds of $K_n$ 
 indicated by Theorem \ref{theorem:kcon}; 
 i.e., to
\begin{align}
    K_n= \left \lceil \frac{\log n +(k-2)\log \log n}{1-\mu}
\right \rceil. \label{eq:kcon_figthresholdkn}
\end{align}
It is evident that the probability of $k$-connectivity increases sharply from 0 to 1 within a small neighborhood of $K_n$ defined in (\ref{eq:kcon_figthresholdkn}). The last plot in Figure~\ref{fig:kcon}  shows the largest value of $k$ for which the network is $k$-connected in at least 990 out of 1000 realizations for a given $\mu$ and $K_n$. From this plot, we see that to achieve a desired level of reliable connectivity with a probability of at least 99\%, a network designer can trade-off a smaller $\K$ for a larger value of $1-\mu$ and vice versa. For instance, if the goal is to design a secure network of 1000 nodes which is 3-connected with probability 0.99, 
this can be achieved by setting the parameters $(\K,\mu)$ close to $(15,0.1)$, or $(20, 0.3)$ or $(30,0.5)$. 
\section{Main Results: The Giant Component}
It is known from \cite{eletreby2019ISIT} that $\hh$ is connected whp {\em only if} $K_n = \omega(1)$. 
A natural question is then to ask what would happen if $\K$ is {\em bounded}, i.e, when $\K=\OO{(1)}$.
It was shown, again in \cite{eletreby2019ISIT}, that $\hh$ has a positive probability of being {\em not} connected in that case. 
Thus, it is of interest to analyze whether the network has a connected sub-network containing a {\em large} number of nodes, or  it consists merely of {\em small} sub-networks isolated from each other. To answer this question, we formally define connected components and then state our main result characterizing the size of the largest connected component of $\hh$ when $K_n = \OO(1)$.

\begin{definition}[Connected Components]
Nodes $v_1$ and $v_2 \in \nodes$ are said to be {\em connected} if there exists a path of edges connecting them. The connectivity of a pair of nodes forms an equivalence relation on the set of nodes. Consequently, there is a partition of the set of nodes $\nodes$ into non-empty sets  $C_1, C_2, \ldots, C_m$ (referred to as connected components) such that two vertices $v_1$ and $v_2$ are connected if and only if $\exists i \in \{1, \ldots, m\}$ for which $v_1, v_2 \in C_i$; see \cite[p.~13]{bondy1976graph}.
\label{def:concomp}
\end{definition}{}

\par In light of the above definition, a graph is connected if it consists of only one connected component.  In all other cases, the graph is {\em not} connected and has at least two connected components that have no edges in between. It is of interest to analyze the fraction of the nodes contained in the {\em largest} connected component as the number of nodes grows. In particular, 
a graph  with $n$ nodes is said to have a {\em giant} component if its largest connected component is of size  $\Omega(n)$.

Let $\cmax$ denote the set of nodes in the largest connected component of $\hh$. Our main results, presented below, show that $|\cmax| = n - O(1)$ whp. Namely, whp, $\hh$ has a giant component that contains {\em all but finitely many} of the nodes. 
First, we show that the probability of at least $M$ nodes being {\em outside} of $\cmax$ decays exponentially fast with $M$. 
\begin{theorem}
{\sl 
For the inhomogeneous random graph $\hh$ with $K_n \geq 2 ~\forall n$ and  
$K_n =O(1)$ we have for each $M = 1, 2, \ldots$ that 
\begin{align}
&  \pr \left[ |\cmax| \leq n-  M\right] \nonumber \\
 &\leq \frac{\exp\{-M\left(\kk-1\right)(1-\oo(1))\}}{1-{\exp\{-\left(\kk-1\right)(1-\oo(1))\}}} + \oo(1) .\label{eq:gc_theorem}
\end{align}
}  \label{theorem:gc}
\end{theorem}
The proof of Theorem~\ref{theorem:gc} relies on showing the improbability of existence of \emph{cuts} of size in the range $[M, n-M]$ as described in Section~\ref{sec:proof}. This approach is inspired by the technique used 
in \cite{MeiPanconesiRadhakrishnan2008} and differs from the branching process technique typically employed in the random graph literature, e.g., in the case of Erd\H{o}s-R\'enyi graphs \cite[Ch.~4]{van2016random}.

\begin{corollary}
{\sl For the inhomogeneous random graph $\hh$ with $K_n \geq 2 ~\forall n$ and  
$K_n =O(1)$ we have 
\begin{align}{}
|\cmax| = n - O(1) \ \ {\rm whp}.
\label{eq:gc_xn_BigO}
\end{align}
}
\label{cor:gc_2_classes}
\end{corollary}
\vspace{-5mm}
\myproof 
Consider an arbitrary sequence $x_n=\omega(1)$. Substituting $M$ with $x_n$ in (\ref{eq:gc_theorem}), we readily see that 
\begin{align}
  \limit  \pr \left[  n- |\cmax|  \leq  x_n\right]=1.
\end{align}{}
Namely, we have 
\begin{align}
n - |\cmax| \leq x_n \ {\rm whp~for~any} \ \ x_n =\omega(1).
\label{eq:gc_xn_whp}
\end{align}
This is equivalent to the number of nodes ($n - |\cmax|$) outside the largest connected component being  {\em bounded}, i.e., $O(1)$, with high probability. This fact is sometimes stated using the probabilistic big-O notation, $O_p$. A random sequence $f_n = O_p(1)$ if for any $\varepsilon>0$ there exists finite integers $M(\varepsilon)$ and $n(\varepsilon)$ such that
$\pr[f_n > M(\varepsilon)] < \varepsilon$ for all $n \geq n(\varepsilon)$.
In fact, we see from \cite[Lemma~3]{janson2011probability} that (\ref{eq:gc_xn_whp}) is equivalent to having
$  n - |\cmax| = O_p(1)$
Here, we equivalently state this as 
\begin{align}{}
n - |\cmax| = O(1) \ \ {\rm whp}, \nonumber
\end{align}
giving readily (\ref{eq:gc_xn_BigO}).
\myendpf
Corollary~\ref{cor:gc_2_classes} can be extended to inhomogeneous random $K$-out graphs 
with arbitrary number of node types;  see Appendix.

\subsection{Discussion}

Theorem~\ref{theorem:gc} shows that 
for arbitrary $0<\mu<1$ and even with $\K = 2$, the largest connected component in $\hh$ spans $n-\OO(1)$ nodes whp. We expect that especially in resource-constrained environments (e.g., IoT type settings), it will be advantageous to have a 
{\em large} connected component reinforcing the usefulness of the heterogeneous pairwise key predistribution scheme for ensuring secure communications in such applications; see \cite{Haowen_2003,eletrebycdc2018} for other advantages of the (heterogeneous) pairwise scheme.

It is worth emphasizing that the  largest connected component of $\hh$, whose size is given in  (\ref{eq:gc_xn_BigO}), is {\em much larger} than what is strictly required to qualify it as a {\em giant} component; i.e., that $|\cmax| = \Omega(n)$. In fact, for most random graph models, including
Erd\H{o}s-R\'enyi graphs \cite{erdHos1960evolution},
random key graphs
\cite[Theorem 2]{Rybarczyk_2011}, 
studies on the size of the largest connected component are focused on characterizing the behavior of $|C_{\rm max}|/n$ as $n$ gets large; this amounts to studying the {\em fractional} size of the largest connected component. Our result given as (\ref{eq:gc_xn_BigO}) goes beyond looking at the fractional size of the largest component, for which it  gives 
$ \frac{|\cmax|}{n} \to_{p} 1$. 
This is equivalent to having $|\cmax| = n -o(n)$.
However, even having $|\cmax| = n-\oo(n)$ leaves the possibility that as many as $n^{0.99}$ nodes are {\em not} part of the largest connected component. Thus, our result, showing that at most $O(1)$ nodes are outside the largest connected component {\em whp}, is sharper than existing results on the {\em fractional} size of the largest connected component. 

Our result highlights a major difference of inhomogeneous random K-out graphs from classical models such as Erd\H{o}s-R\'enyi (ER) graphs \cite{Bollobas,erdHos1960evolution}. Let $\mathbb{G}(n;p_n)$ denote the ER graph on $n$ nodes and link probability $p_n \in [0,1]$. It is known \cite{erdHos1960evolution},\cite[p.~109, Theorem~5.4]{JansonLuczakRucinski} that with $p=c/n$ and $c>1$ the ER graph has a {\em giant} component of size $\beta n (1+\oo(1))$ whp where $\beta \in (0,1]$ is the solution of $\beta + e^{-\beta c}=1$; if $c<1$, then whp the largest connected component is of size $\OO(\log n)$. With $p=c/n$, the mean node degree in ER graphs equals $c$. To provide an example comparison of the size of the giant component in $\hh$ and ER graphs, let $\K=2$ and $\mu=0.9$. In that case, the mean degree of a node in $\hh$ equals $ (1-\oo(1)) 2 \kk  \approx 2 (0.9+0.1\times2)=2.2$; see Appendix. 
An ER graph with $p=2.2/n$ would have the same mean degree and thus the mean number of edges in both models would {\em match} under these conditions. 
From the above discussion, the largest connected component of the ER graph would be of size $\approx 0.8437n +o(n)$
 whp.  For a network of $5000$ nodes, this corresponds to over 700 nodes being isolated from the largest component. In contrast, Theorem \ref{theorem:gc} shows that the largest connected component of $\hh$ would be much larger. Namely it will be of size $n-O(1)$ whp. 
 This is verified in our numerical experiments in the succeeding section (see Figure~\ref{fig:gc_varymu}), where it is seen that for a network of $5000$ nodes, at most 45 nodes are seen to be outside of the largest connected component  in 100,000 experiments.

 \subsection{Numerical results}

\begin{figure}[!t]
\centering
\includegraphics[scale=0.31]{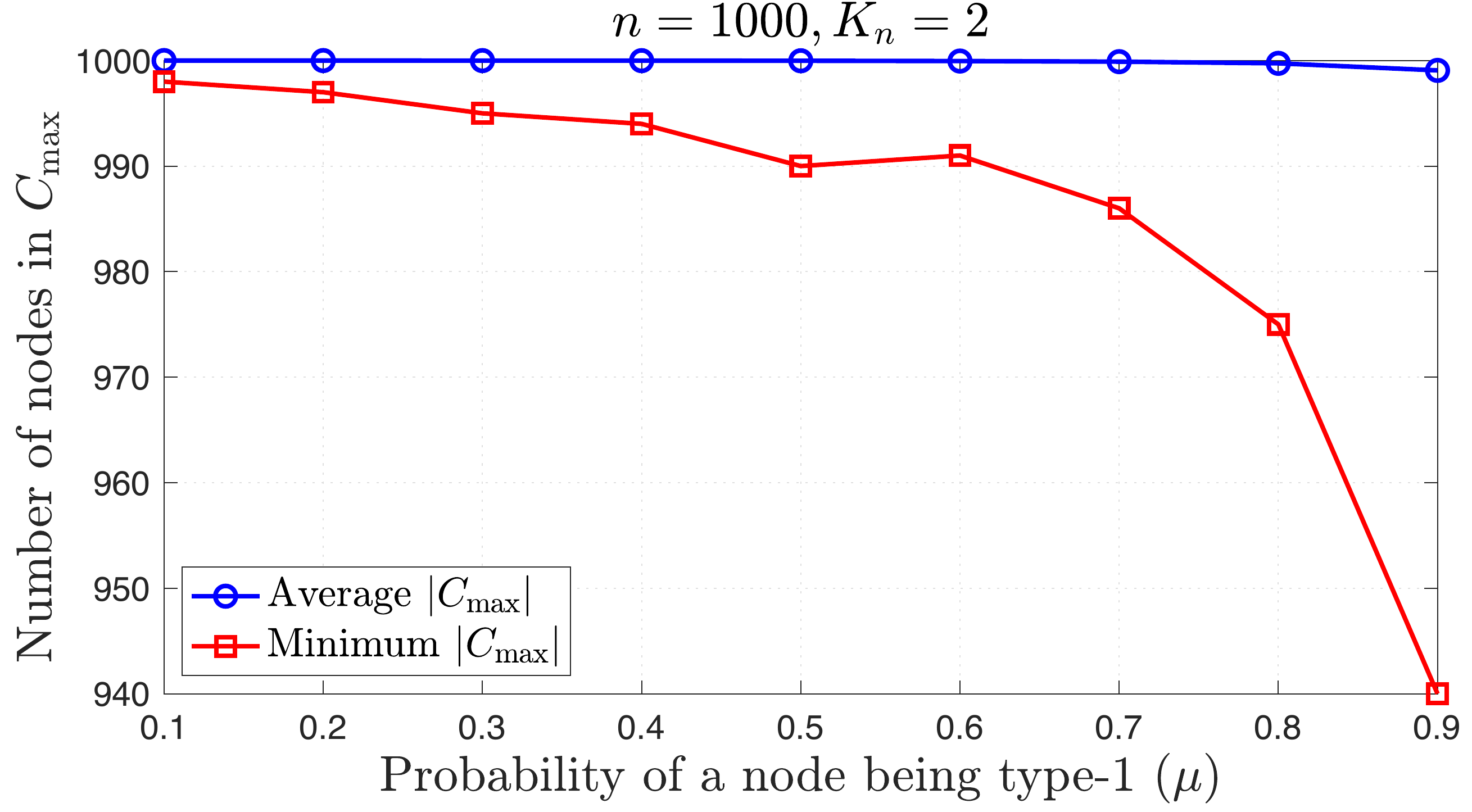}\label{fig:gc1000}
\includegraphics[scale=0.31]{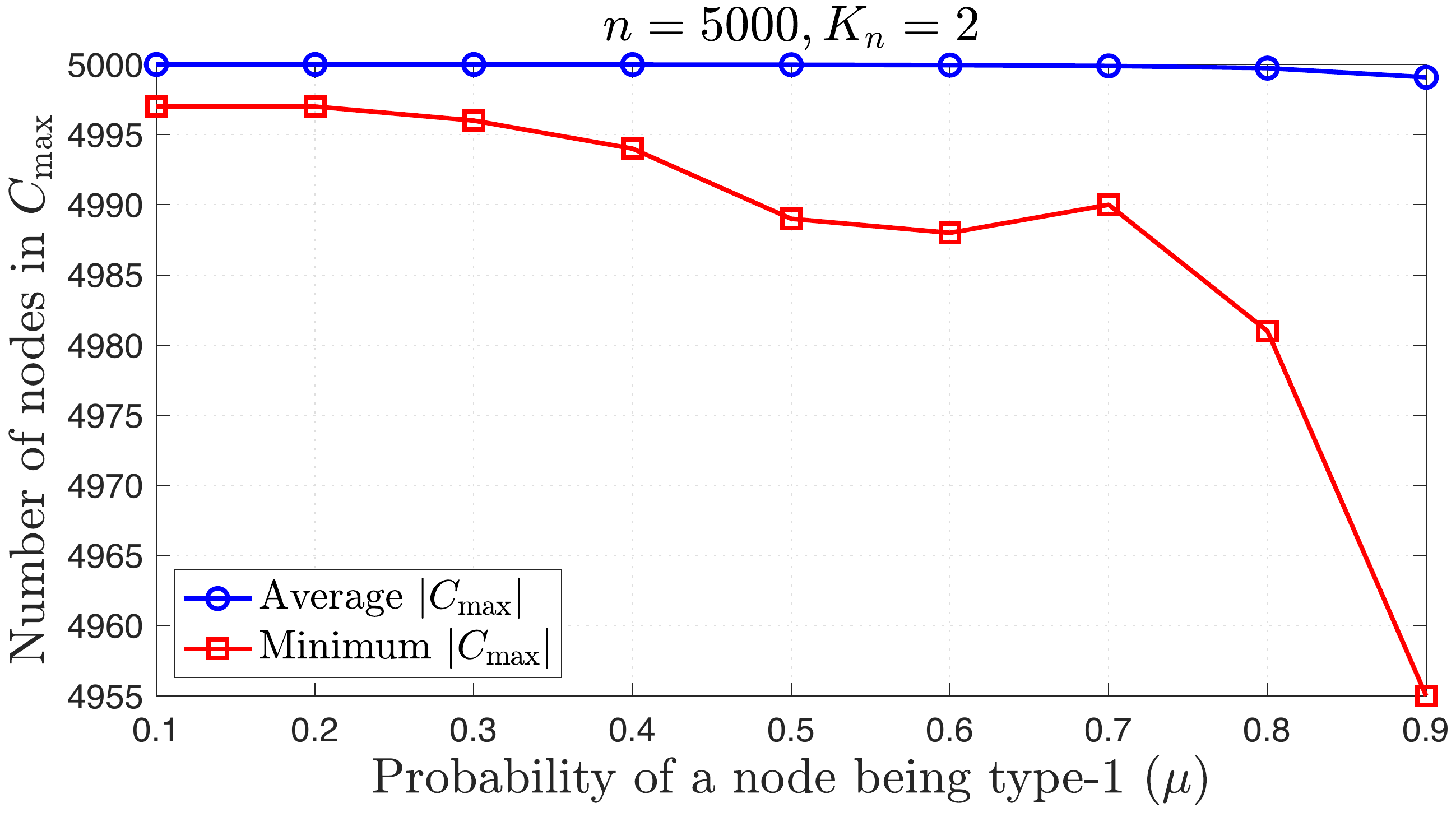}\label{fig:gc5000}
\vspace{-1mm}
\caption{\sl Average  and minimum number of nodes contained in the largest connected component of $~\hh$  with $K_n=2$, $n=1000,5000$ and $\mu \in \{0.1, \dots, 0.9\}$. Even when $\mu=0.9$, setting $K_n=2$ is enough to ensure that almost all of the nodes form a connected component; at most 45 out of 5000 nodes (or, 60 out of 1000 nodes) are seen to be isolated from the giant component across 100,000 experiments. \vspace{-2mm}} 
\label{fig:gc_varymu}
\end{figure}

 \begin{figure}
\centering
\includegraphics[scale=0.32]{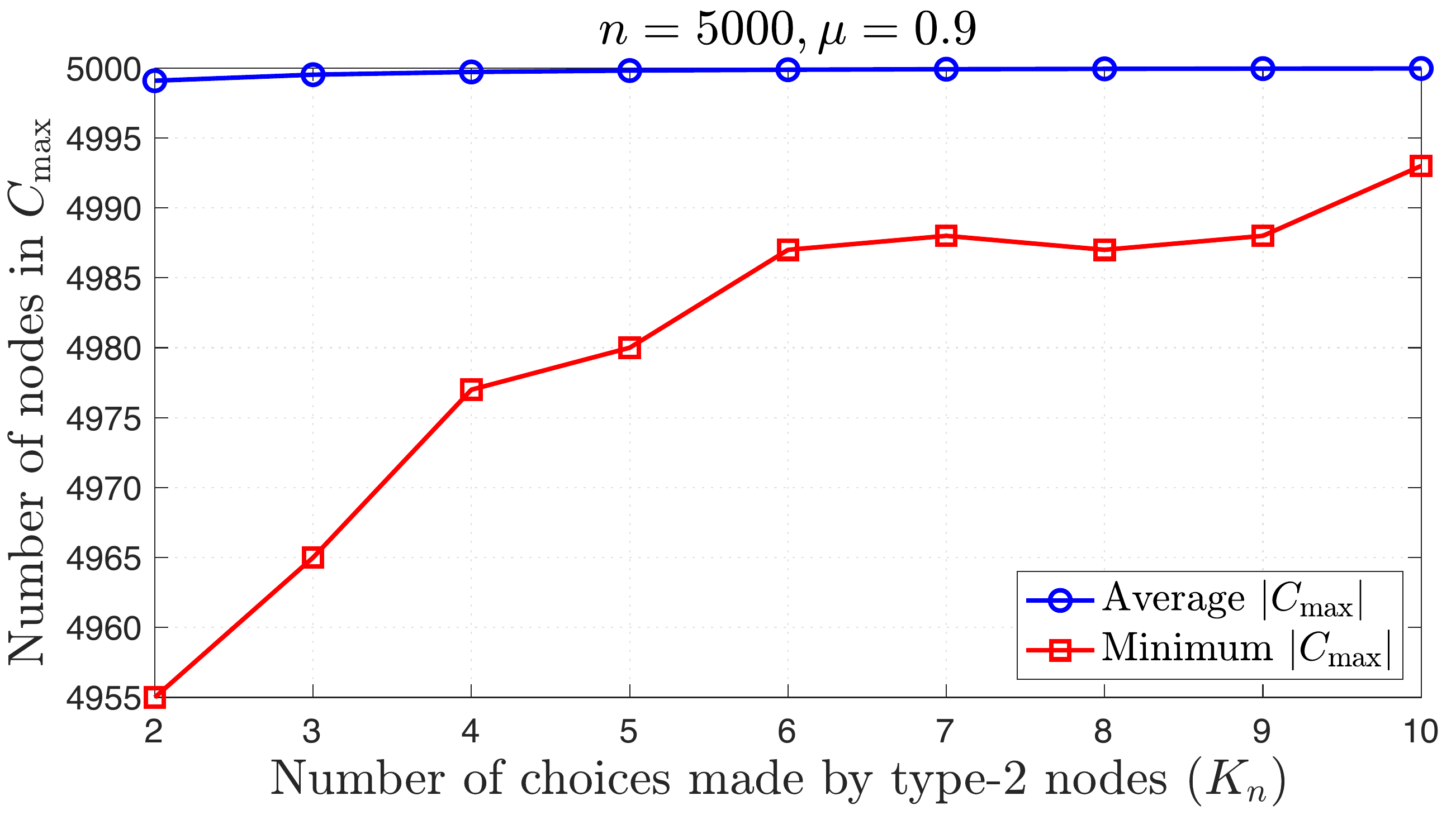} 
\vspace{-3mm}
\caption{\sl Average  and minimum number of nodes contained in the largest connected component of $~\hh$ across 100,000 experiments with $n=5000, \mu=0.9$ and $K_n \in \{2, \dots, 10\}$. \vspace{-2mm}}
\label{fig:gc_varyK}
\end{figure}

Next, we investigate the size of the largest connected component of $\hh$ when the number of nodes is finite through simulations. Recall from Theorem~\ref{theorem:gc} that for $\K\geq 2$ the largest connected component is of size $n-\OO(1)$  whp; i.e, all but finitely many nodes are in the largest connected component. We present empirical studies probing the applicability of this result in the non-asymptotic regime. 
 
 We first explore the impact of
 varying the probability $\mu$ of a node being type-1 nodes 
 on the size of the largest connected component. We generate 100,000 independent realizations of $\hh$ with $\K=2$ for $n=1000$ and  $n=5000$, varying $\mu$ between 0.1 and 0.9 in increments of 0.1. 
We first focus on the {\em minimum} size of the largest component observed in 100,000 experiments. Then we plot the {\em average} size of the largest component, as shown in Figure~\ref{fig:gc_varymu}.  
 We see that even when the probability of a node being type-1
 is as high as 0.9, setting $\K=2$ is enough to have almost all of the nodes to form a connected component. For $n=1000$ and $5000$, at most 60 and 45 nodes, respectively, are found to be outside of the largest connected component. The observation that the number of nodes outside the largest connected component does not scale with $n$ is consistent with Theorem~\ref{theorem:gc} and (\ref{eq:gc_xn_BigO}).
 
 The next set of experiments  probes the impact of varying the number $\K$ of edges pushed by type-2 nodes when $\mu$ is fixed. We generate 100,000 independent realizations of $\hh$ for  $n=5000$ while keeping $\mu$ fixed at $0.9$ and varying $\K$ between 2 and 10 in increments of 1. Increasing $\K$ has an impact similar to decreasing $\mu$ and we see in Figure~\ref{fig:gc_varyK} that both the average and the minimum size of the largest connected component increases nearly monotonically. Given that increasing $\K$ (or, decreasing $\mu$) increases $\kk$ in view of
 (\ref{eq:avg_K}), this observation is consistent with Theorem \ref{theorem:gc} which states that $\pr \left[   n-|\cmax| > M  \right]$ decays to zero exponentially with $(\kk - 1)M$. 

\section{Outline of Proof of Theorem~\ref{theorem:kcon}}
\label{sec:kcon_outline}
In this section, we outline the high-level steps of the proof of Theorem~\ref{theorem:kcon} and present  results that reduce the proof to establishing Proposition~\ref{prop:kcon_thm1} given at the end of this section. The proof of Proposition~\ref{prop:kcon_thm1} is given in the Appendix.

The heterogeneity of node types makes $\hh$ a complicated model and the proofs involve techniques that are different from those used for the homogeneous K-out random graph \cite{FennerFrieze1982}, \cite{Yagan2013Pairwise}.
For instance, certain correlations that are known to exist amongst events of interest in homogeneous K-out graphs do not necessarily hold in $\hh$. The simplest example is the events that encode the existence of an edge between nodes $v_x$ and $v_y$ ($E{xy}$) and $v_x$ and $v_z$
($E{xz}$). These are known to be negatively correlated in the homogeneous K-out graph \cite{yagan2012modeling}. 
However, in $\hh$, if we condition on $E{xy}$,  two competing factors come into play. Under $E{xy}$ it becomes more likely that $v_x$ is type-2 meaning that it is now {\em more} likely to be connected to $v_z$. However, $E{xy}$ also means that either $v_y$ picked $v_x$ or $v_x$ picked $v_y$. The latter event means that $v_x$ already used one of its choices and thus became {\em less} likely to be connected to another node $v_z$.
Due to these difficulties, some of the key bounds used in our proof are obtained via conditioning on the types of nodes. 
\subsection{Proving the zero-law: From minimum node degree to $k$-connectivity}

Consider an inhomogeneous random K-out graph $\hh$ as given in the statement of Theorem \ref{theorem:kcon}
with the sequence $\g: \N_0\rightarrow \R$  defined through (\ref{eq:kcon_scaling}) for $k\geq 2$. Let $\delta$ denote the {\em minimum} node degree in $\hh$, i.e., 
$
\delta := \min_{i=1,\ldots, n}\{\textrm{deg}(v_i)\},
$ 
with $\textrm{deg}(v_i)$ denoting the number of edges incident on vertex $v_i$. A zero-one law for the minimum node degree of $\hh$ was established in  \cite[Theorem 1]{extvs}. Namely, it was shown for all $k=2,3,\ldots$ that for $\gamma_n$ defined through (\ref{eq:kcon_scaling}),
\begin{align}
  \limit \pr\left[
\delta \geq k
  \hspace{1mm}\right]
  = \begin{cases}
                                  1 & \textrm{if } \limit \g_n =+\infty, \\ 
                                  0 & \textrm{if } \limit \g_n =-\infty. \\
  \end{cases}
  \label{eq:min_node_degree}
\end{align}

Let $\kappa_v$ denote the {\em minimum} number of vertices that need to be removed from $\hh$ to make it {\em not} connected. As before, we say that $\hh$ is $k$-connected if $\kappa_v \geq k$. We always have
\begin{align}
    \kappa_v \leq \delta \label{eq:kcon_mnd_basic}
\end{align}
since removing all neighbors of a node with degree $\delta$ would render the node {\em isolated}, making the graph disconnected. Thus, for all $k=1, 2, \ldots$, it holds that 
    $[\kappa_v \geq k] \subseteq [ \delta \geq k]$,
which gives
\begin{align}
    \pr[\kappa_v \geq k] \leq \pr[\delta \geq k]. \label{eq:kcon_mnd_zero}
\end{align}
In view (\ref{eq:kcon_mnd_zero}), the zero-law given in (\ref{eq:min_node_degree}) leads to
\begin{align}
    \limit \pr\left[
   \textrm{$\hh$ is $k$-connected}\right] = 0 \quad \textrm{if~~}\limit \g_n =-\infty
\end{align}
establishing the zero-law of Theorem \ref{theorem:kcon}.


\subsection{A sufficient condition for the one-law for $k$-connectivity}
 
The rest of this section is devoted to proving the one-law of Theorem~\ref{theorem:kcon}, namely showing that 
\begin{align}
    \limit \pr\left[
   \textrm{$\hh$ is $k$-connected}\right] = 1 \quad \textrm{if~~}\limit \g_n = +\infty
   \label{eq:to_show_one_law}
\end{align}
From  (\ref{eq:min_node_degree}) we see that
$\pr[\delta \geq k] \to 1$ 
when $\g_n \to +\infty$. To leverage this result, we write
\begin{align}
\pr[\kv \geq k] 
&=\pr[\kv \geq k, ~ \delta \geq k] 
\label{eq:kcon_1l_1} \\
&= \pr[\delta \geq k] - \pr [ \delta \geq k, \kappa_v <k], \nonumber\\
&=  \pr[\delta \geq k] - \pr \left[ \cup_{\ell=0}^{k-1} \{\delta \geq k, \kappa_v = \ell \}\right]
\nonumber \\ 
&\geq \pr[\delta \geq k] - \pr \left[  \cup_{\ell=0}^{k-1} \{\delta > \ell, \kappa_v = \ell \}\right] \nonumber\\
&= \pr[\delta \geq k] -  \sum_{\ell=0}^{k-1} \pr \left[  \delta > \ell, \kappa_v = \ell \right] \label{eq:kcon_suff_cond_eq_sum}
\end{align}
where (\ref{eq:kcon_1l_1}) is a consequence of (\ref{eq:kcon_mnd_basic}).
Using the one-law of (\ref{eq:min_node_degree})
in
(\ref{eq:kcon_suff_cond_eq_sum}), 
we see that the one-law for $k$-connectivity (i.e., (\ref{eq:to_show_one_law})) will follow if we establish that
\begin{align}
&\limit \pr[\delta>\ell,\kv=\ell]=0 \ \nonumber\\ 
&\textrm{ if } \limit{\g_n} =+\infty, \ \ \forall \ \ell=0,1,\ldots,k-1.  \label{eq:kcon_suff_cond_eq}
\end{align}

Conditions in (\ref{eq:kcon_suff_cond_eq}) 
encode the {\em improbability} for $\hh$ to have minimum node degree of at least $\ell + 1$ and yet be disconnected by deletion of a set of $\ell$ nodes. 
In the subsequent sections, we establish (\ref{eq:kcon_suff_cond_eq}) by deriving a tight upper bound on $\pr[\delta>\ell,\kv=\ell]$ which goes to zero as $n$ gets large for each $\ell = 0, 1, \ldots, k-1$. This approach has also proved useful in establishing one-laws for $k$-connectivity in many other random graph models including Erd\H{o}s R\'enyi (ER) graphs \cite[p.~164]{Bollobas}, random key graphs, intersection of ER graphs and homogeneous K-out graphs \cite{yavuz2017k}, etc.



 
\subsection{A reduction step}
In this section we show that while proving the sufficient condition (\ref{eq:kcon_suff_cond_eq}) for $k$-connectivity, we can restrict our analysis to the subclass of sequences $\gamma_n$ defined through (\ref{eq:kcon_scaling}) that scale as $\OO(\log n)$. As the next result shows, the desired one-law for $k$-connectivity (without any constraint on $\gamma_n$) would follow upon establishing it for the constrained scaling.
\begin{lemma}
Consider a scaling $K:\N_0 \rightarrow \N_0$ and $\mu$ such that $0<\mu<1$. With an integer $k\geq2$ let the sequence  $\gamma_n$ be defined through (\ref{eq:kcon_scaling}). \\
If it holds that
\begin{align}
&\limit{\g_n}=+\infty \  \textrm{~and~}  \ \g_n = \OO (\log n) 
\nonumber \\ \nonumber
& \implies 
 \limit  \pr[\hh \textrm{ is $k$-connected } ] =1
\end{align}
then the following implication also holds
\[ 
\limit{\g_n}=+\infty \implies 
 \limit  \pr[\hh \textrm{ is $k$-connected } ] =1.
\]
\label{lem:coupling_summary}
\end{lemma}{}
Lemma \ref{lem:coupling_summary} states that we can assume  the condition $\gamma_n=\OO(\log n)$ in proving the one-law for $k$-connectivity in Theorem~\ref{theorem:kcon} without any loss of generality. 
Its proof passes through showing that for any scaling
$K: \N_0 \rightarrow \N_0$ satisfying ${\g_n} \to +\infty$, we can construct an auxiliary scaling $\tilde{K}: \N_0 \rightarrow \N_0$ such that i) the corresponding sequence $\tilde{\gamma}_n$ satisfies both
${\tilde{\gamma}_n} \to +\infty$ and $\tilde{\gamma}_n=O(\log n)$; and ii) $\tilde{K}_n \leq K_n$ for all $n =2, 3, \ldots$. In view of the second fact, we then provide a formal coupling argument showing that 
\begin{align}
    &\pr[\mathbb{H}(n;\mu, \tilde{K}_n) \textrm{ is $k$-connected }] \nonumber \\ &\leq  \pr[\hh \textrm{ is $k$-connected } ].
\end{align}{}

A proof of Lemma~\ref{lem:coupling_summary} with all details is given in the Appendix.

We find it convenient to introduce the notion of {\em admissible} scaling to characterize mappings that satisfy the additional condition $\gamma_n = O(\log n)$. Recall that
any mapping $K:\N_0 \rightarrow \N_0$ satisfying the conditions 
\begin{align}
2 \leq K_n < n, \quad  n=2, 3, \ldots,
\label{eq:scaling_condition}
\end{align} 
as a \emph{scaling}.

\begin{definition}[Admissible Scaling]\label{def:kcon_admissible}
A mapping $K:\N_0 \rightarrow \N_0$ is said to be an \emph{admissible scaling} if  (\ref{eq:scaling_condition}) holds and the sequence $\gamma_n$ defined through (\ref{eq:kcon_scaling}) satisfies $\gamma_n = \OO (\log n)$.
\end{definition}{}

It is now clear that the proof of Theorem~\ref{theorem:kcon} will be completed if we establish (\ref{eq:kcon_suff_cond_eq}) for any {\em admissible} scaling. 
This result is presented separately for convenience as follows. 
\begin{proposition}
\label{prop:kcon_thm1}
{ 
With $0<\mu<1$ and an integer $k \geq 2$, consider an admissible scaling $K: \N_0 \rightarrow \N_0$ with the sequence $\gamma_n$ defined through (\ref{eq:kcon_scaling}).
We have
\begin{align}
&\limit \pr[\delta>\ell,\kv=\ell]=0 \nonumber \\
&\textrm{ if } \limit{\g_n} =+\infty, \ \ \forall \ \ell=0,1,\ldots,k-1.  
\nonumber 
\end{align}
\label{prop:kcon_constrained}
}
\end{proposition}{}
The main benefit of being able to restrict the discussion to admissible scalings is to have 
$
K_n = \Theta(\log n)
$
when $\gamma_n \to \infty$ in view of (\ref{eq:kcon_scaling}). This condition will prove useful in  bounding $\pr[\delta>\ell,\kv=\ell]$ efficiently in several places. 



\section{Outline of Proof of Theorem~\ref{theorem:gc}}
\label{sec:proof}

Recall that $\cmax$ denotes the largest connected component of $\hh$. In this section,  we will give an outline of the proof of Theorem~\ref{theorem:gc}. All details can be found in the Appendix. 
The proof of this result goes through a sequence of intermediate steps. We start  by  defining a {\em cut} as a subset of nodes that is isolated from the rest of the graph.

\begin{definition}[Cut]
\cite[Definition 6.3]{MeiPanconesiRadhakrishnan2008}
Consider a graph $\mathcal{G}$ with the node set $\nodes$. A \emph{cut} is defined as a non-empty subset $S \subset \nodes$ of nodes 
that is {\em isolated} from the rest of the graph. Namely,  $S \subset \nodes$ is a cut if there is no   edge between $S$ and $S\comp=\nodes \setminus S$.  
\label{def:cut}
\end{definition}{}
It is clear from Definition~\ref{def:cut} that if $S$ is a cut, then so is $S\comp$.
It is important to note the distinction between a \emph{cut} as defined above and the notion of a \emph{connected component} given in Definition~\ref{def:concomp}. A connected component is isolated from the rest of the nodes by Definition~\ref{def:concomp} and therefore it is also a cut. However, nodes within a cut may not be connected meaning that not every cut is a connected component. 

Let $\mathcal{E}_n (\mu,K_n; S)$ denote the event that  $S \subset \nodes$ is a cut in $\hh$ as per Definition~\ref{def:cut}. Event $\mathcal{E}_n (\mu,K_n; S)$ occurs if no nodes in $S$ pick neighbors in $S\comp$ and no nodes in $S$ pick neighbors in $S\comp$. Thus, we have
\begin{align}
\mathcal{E}_n (\mu,K_n; S) =
\bigcap_{i \in S} \bigcap_{j \in S^c}
\left(
\left \{ i \not \in \Gamma_{n,j} \right \}
\cap 
\left \{ j \notin \Gamma_{n,i} \right \}
\right). \nonumber
\end{align}

Let $\mathcal{Z}(x_n;\mu,K_n)$ denote the event that $\hh$ has no cut $S \subset \nodes$ with size  $x_n \leq |S| \leq n-x_n$ where  $x:\N_0 \rightarrow  \N_0$ is a sequence such that $x_n \leq {n}/{2} \ \forall n$. 
In other words, $\mathcal{Z}(x_n;\mu,K_n)$ is the event that there are no cuts in $\hh$ whose size falls in the range $[x_n, n-x_n]$. Since if $S$ is a cut, then so is $S^{c}$ (i.e., if there is a cut of size $m$ then there must be a cut of size $n-m$), we see that  
\begin{align}
\mathcal{Z}(x_n;\mu,K_n) & = \bigcap_{S \in \mathcal{P}_n: ~x_n\leq  |S| \leq \lfloor \frac{n}{2} \rfloor}  \left(\mathcal{E}_n({\mu},{K}_n; S)\right)\comp,
\nonumber 
\end{align}
where $\mathcal{P}_n$ is the collection of all non-empty  subsets of $\nodes$.  Next, we present an upper bound on $\pr\left[\left(\mathcal{Z}(M;\mu,K_n)\right)\comp\right]$, 
i.e, the probability that there exists a cut with size in the range $[M,  n-M]$ for $\hh$.
\begin{proposition}
\label{prop:gcproofk3}
Consider a scaling $K: \N_0 \rightarrow \N_0$ such that $\K \geq 2 $  $\forall n$ and $K_n = O(1)$, 
and $\mu \in (0,1)$. It holds that 
\begin{align}
 & \pr\left[\left(\mathcal{Z}(M;\mu,K_n)\right)\comp\right]\nonumber \\
 &\leq \frac{\exp\{-M\left(\kk-1\right)(1-\oo(1))\}}{1-{\exp\{-\left(\kk-1\right)(1-\oo(1))\}}} + \oo(1) .\label{eq:gcpropsum}
\end{align}{}
\end{proposition}{}
The proof of Proposition~\ref{prop:gcproofk3} is given in the Appendix.

The following Lemma establishes the relevance of the event $ \mathcal{Z}(x_n;\mu,K_n)$ in obtaining a lower bound for the size of the largest connected component.  
\begin{lemma}
\label{lem:gc_sum}
For any sequence $x:\N_0 \rightarrow \N_0$ such that $x_n \leq \lfloor n/3 \rfloor $ for all $n$, we have
\begin{align*}
 \mathcal{Z}(x_n;\mu,K_n) \implies    |\cmax| > n -x_n  . 
\end{align*}{}
\end{lemma}{}
\vspace{-3mm}
\myproof
Assume that 
$\mathcal{Z}(x_n;\mu,K_n)$
takes place, i.e.,  
 there is no cut in $\hh$ of size in the range $\left[x_n,n-x_n\right]$. Since a connected component is also a cut, this also means that 
 there is no connected component of size in the range $[x_n,n-x_n]$. Since every graph has at least one connected component, it either holds that the largest one has size $|\cmax| > n-x_n$, or that 
 $|\cmax| < x_n$. 
We now show that it must be the case that  $|\cmax|> n-x_n$ under the assumption that $x_n \leq  n/3$. 
Assume towards a contradiction that $|\cmax|<x_n$
meaning that 
the size of each connected component is less than $x_n$. Note that the union of any set of connected components is either a cut, or it spans the entire network. If no cut exists with size in the range $[x_n, n-x_n]$, then the union of any set of connected components should also have a size outside of $[x_n, n-x_n]$. Also, the union of all connected components has size $n$.  Let $C_1, C_2, \ldots, C_{\text{max}}$ denote the set of connected components in increasing size order. Let $m \geq 1$ be the largest integer such that $\sum_{i=1}^m |C_i| < x_n$. Since $|C_{m+1}| < x_n$, we have
\[
x_n \leq \sum_{i=1}^{m+1} |C_i| < x_n+x_n \leq 2n/3 \leq n -x_n.
\]
This means that $\cup_{i=1}^{m+1} C_i$ constitutes a cut with size
in the range $[x_n, n-x_n]$
contradicting the event $\mathcal{Z}(x_n;\mu,K_n)$. 
We thus conclude that if $\mathcal{Z}(x_n;\mu,K_n)$ takes place with $x_n \leq n/3$, then we must have
$|\cmax|> n-x_n$.
\myendpf



\vspace{-2mm}

We now have all the requisite ingredients for establishing Theorem~\ref{theorem:gc}. Substituting $x_n=M, ~ \forall n$ in Lemma~\ref{lem:gc_sum} 
for some finite integer $M$,
we get  
$     \mathcal{Z}(M;\mu,K_n) \implies    |\cmax| > n -M$.
Equivalently, we have
$
     |\cmax| \leq n -M  \implies \mathcal{Z}(M;\mu,K_n) \comp  
$. This gives
\begin{equation}
 \pr\left[{|\cmax| \leq n -M} \right ] \leq \pr\left[{\mathcal{Z}(M;\mu,K_n) \comp}\right]
 \label{eq:gcl1_new}
\end{equation}
and we get (\ref{eq:gc_theorem}) by using 
 Proposition~\ref{prop:gcproofk3} in (\ref{eq:gcl1_new}).

\section{Conclusion}

This paper analyzes the strength of connectivity in inhomogeneous random K-out graphs. In particular, we derive conditions on the network parameters $n,\mu, \K$, which make the graph $k$-connected with high probability. In cases where the parameter $\K$ is constrained to be small, we proved that whenever $\K\geq2$, the largest connected sub-network spans all but finitely many nodes of the network with high probability. Our results complement the existing results on $1$-connectivity of inhomogeneous random K-out graphs. An open direction is characterizing the asymptotic size of the largest connected component of the homogeneous K-out random graph when $K=1$. It would also be of great interest to analyze $k$-connectivity of inhomogeneous random K-out graphs with $r>2$ node types and arbitrary parameters $K_1, K_2, \ldots, K_r$ associated with each node type. Finally, it would be interesting to pursue further applications of inhomogeneous K-out graphs in the context of cryptographic payment channel networks.


\section*{Acknowledgment}
This work has been supported in part by the
National Science Foundation through grant CCF \#1617934.

%

\section*{\Large  \bf Appendix A ($k$-connectivity)} 
\vspace{15 pt}
\setcounter{subsection}{0}
\section*{\textbf{\ \large Appendix A.1:  Key Bounds}}

\label{sec:proofkcon2}
In this section, we present several steps to obtain an upper bound on $\pr[\delta>\ell,\kv=\ell]$.
The proof of Proposition~\ref{prop:kcon_thm1} is completed in Appendix A.4 as we show that the obtained upper bound approaches zero as $n$ gets large.

\subsection{\bf Towards an upper bound on $\pr[\delta>\ell,\kv=\ell]$}

Recall that $\hh$ denotes the inhomogeneous random K-out graph on $n$ nodes with $\mathcal{N}$ being the set of node labels $\{1,2,\ldots,n\}$. For any $\ V \subseteq \mathcal{N}$, let $\iii (V)$ denote the induced sub-graph obtained by restricting $\hh$ to the subset of nodes in $V$ i.e., $\iii (V)$ is a graph with vertex set $\{v_i: i \in V\}$ and edge set given by the subset of edges of $\hh$ which have both endpoints in $V$. Put differently, $\iii (V)$ is the graph induced from $\hh$ upon deletion of nodes in $V\comp$, where $V \subseteq \nodes$ and $V\comp = \nodes \setminus V$.

\begin{figure}[t]
\centering\includegraphics[scale=0.16]{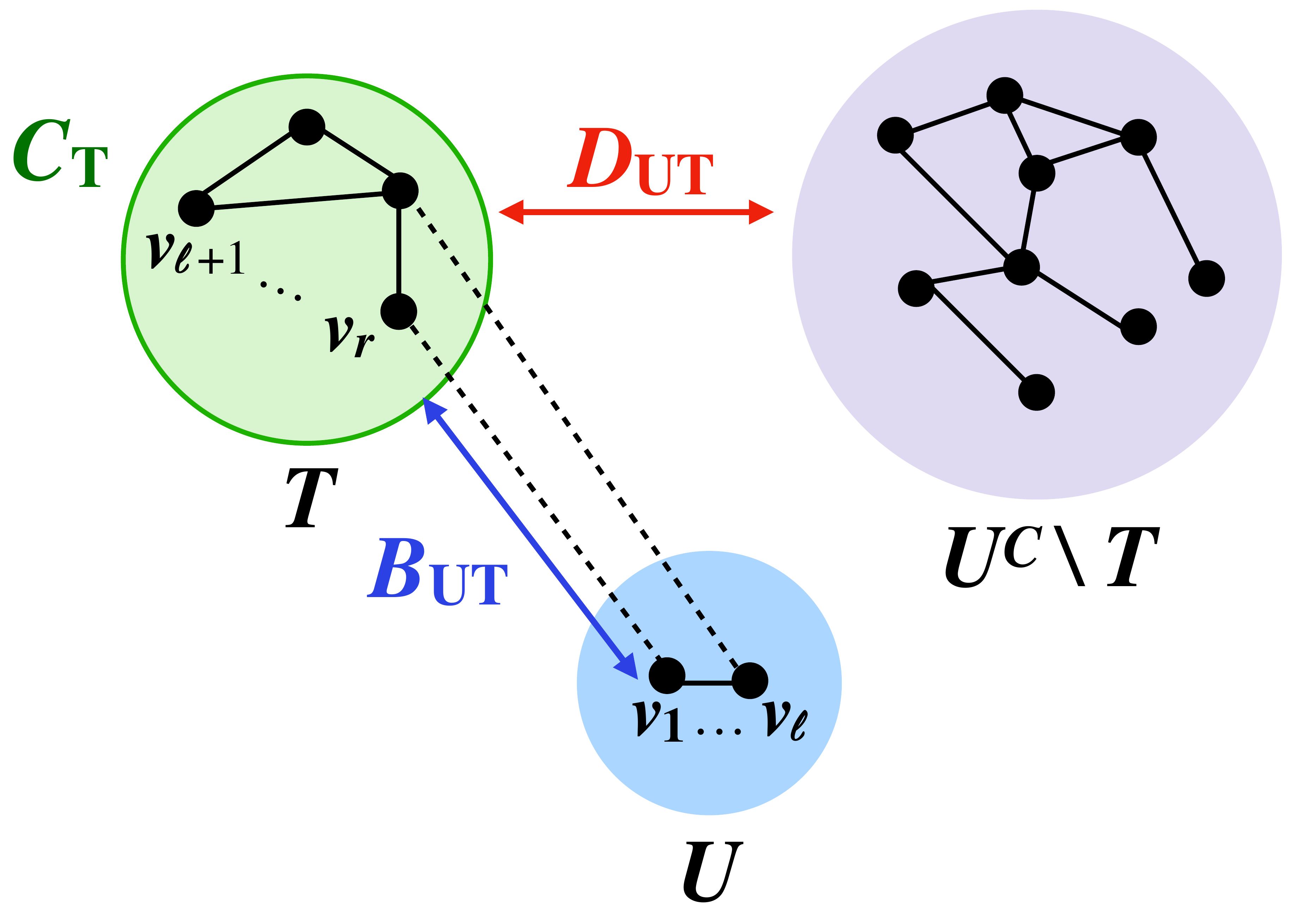}
\vspace{-2mm}
\caption{ A necessary condition for the event $[\kappa_v=\ell,~ \delta > \ell]$ to take place is the occurrence of the event $\aut$ defined as the intersection of $\ct, \dut$ and $\but$.
}\label{fig:2} 
\vspace{-4mm}
\end{figure}

We will establish an upper bound on $\pr[\kv=\ell, \delta>\ell]$ where $0 \leq \ell \leq k-1$ by considering events which follow under  $\{ \delta>\ell, \kv=\ell\}$;  i.e., that the graph $\hh$ has minimum degree strictly greater than $ \ell$ and yet it can be made disconnected by deleting a set of $\ell$ nodes. 
If $\kv=\ell$, then by definition  there exists a vertex cut $U \subset \nodes$ containing $\ell$ nodes such that the deletion of $U$ disconnects $\hh$. More precisely, there must exist $U \subset \nodes$ with $|U|=\ell$ and $T \subset U^c$ such that
$\hh$ is $(U,T)$-{\em separable};  i.e.,  deleting all nodes in $U$ renders $\hh$ disconnected into a subgraph on vertices $T$ and a subgraph on vertices $U^c/T$ with no edges in between the two subgraphs (see Figure \ref{fig:2}). Without any loss of generality, we let $T$ denote the smaller of the sets $T$ and $U^c/T$. Thus, we have  $|T|\leq \lfloor \frac{n-\ell}{2} \rfloor$.

We can further make the following observations regarding the sets $U$, $T$, and $U^c/T$ under the event  $\{\kv=\ell, \delta>\ell\}$:
\begin{enumerate}
    \item Consider an arbitrary node $v \in T$. Since there are no edges between $T$ and $(U\comp)/T$, all edges incident on $v$ must have their endpoints in the set $U \cup T\setminus \{v\}$. Further, since $\delta > \ell$, there are at least $\ell+1$ edges from $v$ to $U \cup T\setminus \{v\}$. Noting that $|U|=\ell$, there must be an edge with between $v$ and $T\setminus \{v\}$ and thus $|T|\geq2$.
    \item For a given vertex cut $U$ (whose deletion makes $\hh$ disconnected) there might be more than one set $T \subset U^{c}$ for which $\hh$ is $(U,T)$-separable. Thus, it suffices to identify the smallest set $T$ for which $\iii(T)$ is {\em connected}. In other words, if $\kappa_v = \ell$ and $\delta> \ell$, then $\hh$ should be $(U,T)$-separable for some $U \subset \nodes$ with $|U|=\ell$ and $T \subset U^c$ with $2 \leq |T| \leq \lfloor \frac{n-\ell}{2} \rfloor$ such that $\iii(V)$ is connected.
    Let $\ct$ denote the event that $\iii(T)$ is connected.
    \item Each node in $U$ must have a neighbor in $T$. Assume towards a contradiction that there exists a node $u$ in $U$ which does not have a neighbor in $T$. This means that $U\setminus \{u\}$ forms a vertex cut of size $\ell-1$ in $\hh$ which contradicts the fact that  $\kappa_v \geq \ell$.
    Let $\but$ denote the event that each node in $U$ has a neighbor in $T$.
\end{enumerate}{}
The above observations are depicted in Figure~\ref{fig:2} and summarized below. For $\hh$, If $\delta>\ell$ and $\kv=\ell$ then $\exists U, T \subset \nodes$ such that $|U|=\ell$ and $2 \leq |T| \leq \lfloor \frac{n-\ell}{2} \rfloor$ and the following events occur.
\begin{enumerate}
    \item[1.] $\ct$: $\iii (T)$ is connected,
    \item[2.] $\but$: All nodes in $U$ have a neighbor in the set $T$,
    \item[3.] $\dut$: $T$ is {\em isolated} in $\iii(U\comp)$; i.e., there are no edges in $\hh$ between nodes in $T$ and nodes in $U^c/T$.
\end{enumerate}{}
Let $\aut$ be defined as the intersection of these three events; i.e., 
\begin{align}
   \aut:=\ct \cap  \but \cap \dut .  \nonumber
\end{align}
Let $[\mathcal{Z}]_{r}$ denote the collection of subsets of set $\mathcal{Z}$ with exactly $r$ elements.
From the preceding arguments, we see that the following inclusion holds.
\begin{align}
    \{\delta > \ell, \kappa_v = \ell \} \subseteq \cup_{U \in [\nodes]_\ell, T \subset U\comp, 2 \leq |T| \leq \lfloor \frac{n-\ell}{2} \rfloor} \aut \nonumber.
\end{align}
Using a union bound, we get
\begin{align}
    \pr [\delta > \ell, \kappa_v = \ell ] &\leq \sum_{U \in [\mathcal{N}]_\ell, T \subseteq U\comp, 2 \leq |T| \leq \lfloor \frac{n-\ell}{2} \rfloor} \pr[\aut],\nonumber\\
    &= \sum_{r=2}^{\lfloor \frac{n-\ell}{2} \rfloor}
    \sum_{U \in [\mathcal{N}]_\ell, T \in [U\comp]_r  } \pr[\aut],\label{eq:kcon_suff_cond_aut}
\end{align}
For $r=1,\ldots, n-\ell-1$, let $\mathcal{A}_{\{1,\dots,\ell\},\{\ell+1, \dots, \ell+r\}}$, $\mathcal{C}_{\{\ell+1, \dots, \ell+r\}}$, $\mathcal{B}_{\{1,\dots,\ell\},\{\ell+1, \dots, \ell+r\}}$, and $\mathcal{D}_{\{1,\dots,\ell\},\{\ell+1, \dots, \ell+r\}}$ be denoted by $\alr$, $\clr$, $\blr$ and $\dlr$ respectively.
The number of subsets of $\nodes$ of size $\ell$ is ${n \choose \ell}$ and the number of subsets of $\nodes \setminus U$ of size $r$  is ${n-\ell \choose r}$.
Under the enforced assumptions, the exchangeability of the vertex labels yields
\begin{align}
    \pr[\aut]=\pr[\alr],\ \ \ U \in [\mathcal{N}]_\ell, \ T \in [U\comp]_r 
    \label{eq:kcon_suff_cond_exch}
\end{align}
Substituting (\ref{eq:kcon_suff_cond_exch}) in (\ref{eq:kcon_suff_cond_aut}),  we thus get
\begin{align}
    \pr [\delta > \ell, \kappa_v = \ell ] &\leq 
    \sum_{r=2}^{\lfloor \frac{n-\ell}{2} \rfloor}
    {n \choose \ell} {n-\ell \choose r}  \pr[\alr].  \label{eq:kcon_suff_cond_transition}
\end{align}
Combining (\ref{eq:kcon_suff_cond_eq}) and (\ref{eq:kcon_suff_cond_transition}), we observe that to establish the one-law for $k$-connectivity, it suffices to show that the following holds,  
\begin{align}
    \limit  \sum_{r=2}^{\lfloor \frac{n-\ell}{2} \rfloor}
    {n \choose \ell} {n-\ell \choose r}  \pr[\alr] =0,
   \  \ \ \forall \ell = 0,1,\dots,k-1.
    \label{eq:kcon_suff_cond_new}
\end{align}

\subsection{\bf Observations regarding events $\blr$ and $\clr$}
Recall that $\alr= \clr \cap \blr \cap \dlr$. What makes the proof of Proposition~\ref{prop:kcon_thm1} particularly challenging is the intricate correlations among these events. In what follows, we establish several useful bounds that will pave the way for proving Proposition~\ref{prop:kcon_thm1}. 
Define $S:=U\comp \setminus T$. For the remaining sections, we suppress the indices $(\ell, r)$ and work with the notation and definitions listed in Table~\ref{tab:event_notation}. In particular, we define $\dts$ to be the event that for each $i$ in $T$,  the nodes $\Gamma_{i,n}$ picked by $v_i$ should all belong to $S^c$. We define 
$\dst$ similarly as the event that all  $j$ in $S$ pick their choices  $\Gamma_{j,n}$ from outside of $T$.
For $T$ to be isolated in $\iii(U\comp)$, both $\dts$ and $\dst$ should hold, whence $D=\dst \cap \dts$.
With $t=1,2$, we  define $\type$ as the event that all nodes in $T$ are type $t$. Finally, $E_{ij}$ denotes the event that there exists an edge in $\hh$ between nodes $v_i$ and $v_j$; i.e., $E_{ij} = j \in \Gamma_{n,i} ~
\vee~ i \in \Gamma_{n,j}. $


A key observation towards bounding events $B,C$ is the fact
that for any $S \subseteq \nodes$, the random variables  $\{\ii[E _{ij}]: i, j \in S, ~ i\neq j\}$ are {\em negatively associated} conditionally on the types of nodes in $S$. Negative association of random variables, introduced by Joag-Dev and Proschnan \cite{joag1983negative} is a stronger form of negative correlation, and is formally defined next.

 \begin{table}
 \vspace{-1mm}
   \caption{Notation}
    \vspace{-2mm}
   \label{tab:event_notation}
   \begin{tabular}{c|c}
   \hline
     Event&Definition\\
     \hline
     $B$ & All nodes in $U$ have a neighbor in the set $T$.\\
     $C$ & $\iii (T)$ is connected. \\
     $\dst$ & None of the nodes in $S$ pick nodes in $T$ as their neighbors.\\
     $\dts$ & None of the nodes in $T$ pick nodes in $S$ as their neighbors.\\
     $D$ & $T$ is isolated in $\iii (U\comp)$, thus $D=\dst \cap \dts$. \\
     $\eij$ & Nodes $v_i$ and $v_j$ are adjacent, where $i,j \in \nodes$.\\
     $\type$ & All nodes in $T$ are type-$t$, where $t=1,2$.\\
  \hline
 \end{tabular}
 \end{table}

\begin{definition}[Negative Association of RVs]
The rvs $\{ X_\lambda , \ \lambda \in \Lambda \}$ are  said to
be {\sl negatively associated} if for any non-overlapping subsets
$A$ and $B$ of $\Lambda$ and for any monotone increasing mappings
$\varphi : \mathbb{R}^{|A|} \rightarrow \mathbb{R}$ and $\psi :
\mathbb{R}^{|B|} \rightarrow \mathbb{R}$, the covariance
inequality
\begin{equation}
\E\left[ \varphi( X_A ) \psi( X_B ) \right] \leq \E \left[ \varphi( X_A ) \right ] \E \left[
\psi( X_B ) \right] \label{eq:NegativeAssociationDefn}
\end{equation}
holds whenever the expectations in
(\ref{eq:NegativeAssociationDefn}) are well defined and finite.
\end{definition}
An important observation 
is that the negative association
of the rvs $\{ X_\lambda , \ \lambda \in \Lambda \}$ implies \cite[P2, p. 288]{joag1983negative}  the
inequality
\begin{equation}
\E\left[  \prod_{\lambda \in A} f_\lambda (X_\lambda ) \right] \leq
\prod_{\lambda \in A} \E\left[  f_\lambda (X_\lambda ) \right]
\label{eq:NegativeAssociationConsequence}
\end{equation}
where $A$ is a subset of $\Lambda$ and the collection $\{
f_\lambda , \ \lambda \in A \}$ of  mappings $\mathbb{R}
\rightarrow \mathbb{R}_+$ are all monotone increasing. 

It was shown in \cite{yagan2012modeling} that for  the {\em homogeneous} random K-out graph $\mathbb{H}(n;K)$, the edge assignment variables,
$\{\ii[E _{ij}]: i \neq j ,~~~
i,j =1, \ldots , n\}$ are negatively associated. This follows from the fact that 
\begin{equation}
\{ \ii \left[j \in \Gamma_{n,i} \right], \ j \in \nodes-\{i\} \}
\label{eq:NegativeAssociation2}
\end{equation}
form a collection of negatively associated rvs when the sets $\Gamma_{n,i}$
represent a random sample (without replacement) of size $K$ from
$\nodes-\{i\}$; see \cite[Example 3.2(c)]{joag1983negative} for
details. For the inhomogeneous random K-out graph $\hh$, the situation is more intricate since the size of the random samples $\Gamma_{n,i}$ are themselves {\em random} (takes the value one with probability $\mu$ and $K_n$ with probability $1-\mu$). Nevertheless, if we condition on the types of nodes, then the size of the samples $\Gamma_{n,i}$ become fixed and negative association of the collection of rvs in (\ref{eq:NegativeAssociation2}) would still follow. This overlaps with the notion of {\em conditional} negative association introduced recently in \cite{yuan2010conditional}.  
This observation is presented next.

\begin{lemma}
For the inhomogeneous random K-out graph $\hh$, let $\vec{t}_S$ denote the vector of types for nodes in  $S \subseteq \nodes$. Then, the collection of random variables $\{\ii[E_{ij}], i \neq j, ~~ i,j \in S \}$ are conditionally negatively associated given $\vec{t}_S$. Namely, we have that
\begin{align}
    \E\left[  \prod_{i \neq j,  i,j \in S} f (\ii[E_{ij]} ) ~\bigg|~~ \vec{t}_S \right] \leq
\prod_{i \neq j,  i,j \in S} \E\left[  f (\ii[E_{ij}]) ~\big|~~ \vec{t}_S \right]
\label{eq:conditional_negative_association}
\end{align}
for any monotone increasing function $f$. 
\label{lem:conditional_negative_of_eij}
\end{lemma}

The proof of Lemma \ref{lem:conditional_negative_of_eij} follows from the preceding arguments and  negative association of the random variables in (\ref{eq:NegativeAssociation2}) when the size of the mappings $\ii[j \in \Gamma_{n,i}]$ are fixed given $\vec{t}_S$.

Next, we will make repeated use of the preceding observations and Lemma \ref{lem:conditional_negative_of_eij} to obtain upper bounds on the events $B, C$.
\begin{lemma}
\label{lem:kcon_b_corr}
For $r \in \{2,\dots,n-\ell-1 \}$, we have
$$\pr[B~|~\type] \leq \left(r  \pr[\eij~|~ \text{$v_j$ is type-$t$}] \right)^\ell
\text{ where } t=1, 2.$$
\end{lemma}{}
\textbf{Proof of Lemma~\ref{lem:kcon_b_corr}}
The event $B$ is characterized as 
\begin{align}
B= \bigcap\limits_{i=1}^{\ell}  \bigcup\limits_{j=\ell+1}^{\ell+r} \eij.
\label{eq:characterize_B}
\end{align}
Let $\tu \in \{1,2\}^\ell$ be defined as the vector (of size $\ell$) containing the types of nodes in $U$; $v_1, \ldots, v_{\ell}$; i.e., for each $i=1,\dots, \ell$, the type of node $v_i$ is $(\tu)_i$. Given $\type,\tu$, the types of all nodes  associated with the event $B$ (i.e., $v_1, \ldots, v_{\ell+r}$) are determined. Thus, using Lemma \ref{lem:conditional_negative_of_eij} with $S=\{1, \ldots, \ell+r\}$, we see that
the collection of random variables $\{\ii[\eij], i \neq j, i,j=1,\dots, \ell+r \}$ are conditionally {\em negatively associated} given $\type,\tu$. Furthermore, by
the \lq\lq disjoint monotone aggregation" property of negative association
\cite[p. 35]{dubhashi2009concentration}, the  collection of rvs
$
\left\{ \ii\left[\cup_{j=\ell+1}^{\ell+r} E_{ij}\right], \quad i=1,\ldots, \ell \right\}
$
are also conditionally negatively associated given $\type,\tu$. Conditioning on $\tu$, we thus get 
\begin{align}
    &\pr[B~|~\type]\nonumber \\
    &=\E[~\ii[B]~|~\type] \nonumber \\
     &=\E \left[~\E\left[ \ii [B] ~~\big|~ \type , \tu \right] ~\bigg|~ \type \right]\nonumber\\
     &=\E \left[~\E\left[ \prod\limits_{i=1}^{\ell}  \ii \left[\cup_{j=\ell+1}^{\ell+r} \eij\right] ~\big|~ \type , \tu \right] ~~\bigg|~ \type \right] \nonumber\\
    &\leq \E \left[~\prod\limits_{i=1}^{\ell}\E\left[  \ii \left[\cup_{j=\ell+1}^{\ell+r}\eij\right ]~\big|~ \type , \tu \right] ~~\bigg|~ \type \right]\label{eq:lem:kcon_b_corr_1}\\
     &\leq \E \left[~\prod\limits_{i=1}^{\ell}\E \left[\sum\limits_{j=\ell+1}^{\ell+r}\ii[\eij] ~\big|~ \type , \tu \right] ~~\bigg|~ \type \right]\label{eq:intermediate_union} \\
     &= \E \left[~\prod\limits_{i=1}^{\ell} \sum\limits_{j=\ell+1}^{\ell+r} \pr\left[  \eij ~\big|~ \type , \tu \right] ~~\bigg|~ \type \right]
     \nonumber\\
     &=\E \left[ \prod\limits_{i=1}^{\ell} r \left(1- \left(1-\frac{K_t}{n-1}\right) \left(1- \frac{K_{(\tu)_i}}{n-1}\right)\right) ~~\bigg|~ \type \right] \label{eq:intermediate_eij}\\
     &=r^\ell\E \left[ \prod\limits_{i=1}^{\ell}\left(1- \left(1-\frac{K_t}{n-1}\right) \left(1- \frac{K_{(\tu)_i}}{n-1}\right)\right) \right] \label{eq:lem:kcon_b_corr_2}\\
     &=r^\ell \left(\E \left[ \left(1- \left(1-\frac{K_t}{n-1}\right) \left(1- \frac{K_{(\tu)_i}}{n-1}\right)\right) \right] \right)^\ell
  \label{eq:lem:kcon_b_corr_3} \\
     &=\left(r  \pr[\eij~|~ \text{$v_j$ is type-$t$}] \right)^\ell \nonumber
\end{align}
 where (\ref{eq:lem:kcon_b_corr_1}) follows the conditional negative association of random variables $\{\ii[\eij], i \neq j, i,j=1,\dots,\ell+r \}$ given $\type,\tu$,
 (\ref{eq:intermediate_union}) follows from a union bound,
(\ref{eq:intermediate_eij})  follows from the direct computation of $ \pr[\eij~|~ \type, \tu]$ 
 under given types (similar to (\ref{eq:kcon_meandegree})),
 (\ref{eq:lem:kcon_b_corr_2}) follows from the independence of $\type$ and $\tu$, and
 (\ref{eq:lem:kcon_b_corr_3}) is a consequence of the independence of the random variables $\{(\tu)_i, i=1,\dots, \ell\}$.
\myendpf

The next result shows that conditioning on all nodes in $T$ being type-$2$ provides an upper bound on the events $B, C$.
\begin{lemma}
\label{lem:kcon_coupling_t2}
Given that all nodes in $T$ are of type-2, the occurrence of events $B$ and $C$ become more likely, i.e.,
$$ \pr[B,C] \leq \pr[B,C|\ttwo]. $$
\end{lemma}{}
\textbf{Proof of Lemma~\ref{lem:kcon_coupling_t2}}
\begin{align}
    \pr[B,C] = \pr[B,C ~|~ \ttwo]\pr[\ttwo]+\pr[B,C ~|~ \ttwo \comp]\pr[\ttwo \comp]
    \label{eq:kcon_lem_coupling_t2_1}
\end{align}
Thus, it suffices to show that 
\begin{align}
    \pr[B,C ~|~ \ttwo \comp] \leq \pr[B,C ~|~ \ttwo] \label{eq:kcon_lem_coupling_t2_2}
\end{align}
To see this, we note that both events $B$ and $C$ are monotone increasing under edge addition. In other words, if event $B$ (resp.~$C$) occurs in a given  realization $H_1$ of $\hh$, then it will also  occur in any another realization $H_2 \supseteq H_1$.  Given a graph in which $T$ has $m$ $(1\leq m \leq r)$ type-1 nodes, we can construct another graph by selecting precisely $\K-1$  neighbors uniformly at random from the remaining $n-2$ nodes for each type-1 node. This construction shows that the edge set of the realization of $\hh$ in which there is at least one type-1 node in set $T$ is strictly contained in the corresponding edge set when all nodes in $T$ are type-2. 
Thus, we get (\ref{eq:kcon_lem_coupling_t2_2}) as a consequence of both events $B$ and $C$ being monotone-increasing upon edge addition. 

\myendpf

\begin{lemma}
\label{lem:kcon_tree_bound}
For $r \in \{2,\dots,n-\ell-1 \}$, we have
\begin{align}
 \pr[C~|~\type]&\leq r^{r-2} \pr[\eij~|~ \text{$v_i, v_j \in T$ are type-$t$}] ^{r-1}. \nonumber
 \end{align}
\end{lemma}{}

\textbf{Proof of Lemma~\ref{lem:kcon_tree_bound}}

 $C$ is the event that nodes in $T$ form a connected 
subgraph in $\hh$, i.e., that $\iii(T)$ is connected. This is equivalent to stating that $\hh$ contains a spanning tree on nodes
in $T$. Let $\mathcal{Q}_{T}$ denote the collection of all spanning trees on the vertices in $T$.  Put differently, each $Q \in \mathcal{Q}_{T}$
is a collection of $r-1$ node pairs representing the $r-1$ edges of the tree.
Hence, we have
\begin{align}\label{eq:to_compare_C}
C =\bigcup_{Q \in {\mathcal Q}_{T}} \bigcap_{(j_1,j_2) \in Q} E_{j_1j_2}
\end{align}

From Lemma \ref{lem:conditional_negative_of_eij}, we know that the random variables $\{\ii[\eij], i \neq j, ~ i, j \in T \}$ are conditionally negatively associated given the types of nodes in $T$.  
Also, by Cayley's formula \cite[p.~35]{bondy1976graph}, we know that there are $r^{r-2}$ trees on $r$
vertices, i.e., $| {\mathcal Q}_{T}| = r^{r-2}$. 
Using a union bound in (\ref{eq:to_compare_C}), we then get
\begin{align}
    \pr[C~|~\type] 
    &\leq \sum \limits_{Q \in \mathcal{Q}_T}  \E\left[ \prod_{(j_1,j_2) \in Q} \ii[E_{j_1j_2}] ~\big|~~\type\right] \nonumber \\
   & \leq  \sum \limits_{Q \in \mathcal{Q}_T}  \prod_{(j_1,j_2) \in Q}
    \pr \left[ E_{j_1j_2} ~\big|~~\type\right] 
     \label{eq:intermediate_C_lemma}\\
    &= r^{r-2} \pr[\eij~|~ \text{$v_i, v_j \in T$ are type-$t$}] ^{r-1} 
     \label{eq:intermediate2_C_lemma}
     \end{align}
where (\ref{eq:intermediate_C_lemma}) follows from the conditional negative association of $\{\ii[E{j_1j_2}]: j_1,j_2 \in Q\}$ given the types of all nodes in $T \supset Q$, and (\ref{eq:intermediate2_C_lemma}) follows from exchangeability of events $E_{j_1j_2}$.
\myendpf

\begin{lemma}
\label{lem:kcon_bc_corr}
The events $B$ and $C$ are negatively correlated given the types of nodes in $T$, i.e, 
$$\pr[B,C | \type] \leq \pr[B|\type] \pr[C|\type], \qquad t=1,2.$$
\end{lemma}{}
\textbf{Proof of Lemma~\ref{lem:kcon_bc_corr}}
From (\ref{eq:to_compare_C}), we see that 
\begin{align}
\ii[C] =  1- \prod_{Q \in {\mathcal Q}_{T}} \left(1- \prod_{(j_1,j_2) \in Q}  \ii[E_{j_1j_2}]\right)
\label{eq:to_compare_C_2}
\end{align}
while from (\ref{eq:characterize_B}) we have 
\begin{align}
    \ii \left[B\right] =
    \prod_{i =1}^{\ell}\left(1- \prod_{j = \ell+1 }^{\ell + r} \left(1-\ii[E_{ij}]\right) \right).
    \label{eq:to_compare_C_3}
\end{align}
From Lemma \ref{lem:conditional_negative_of_eij}, we know that the random variables $\{\ii[E_{ij}], i \neq j, ~~ i,j = 1, \ldots, \ell+r \}$ are conditionally negatively associated given the types of all nodes $v_1, \ldots, v_{\ell+r}$. We will leverage this fact to show the conditional negative association of rvs $\ii[B]$ and $\ii[C]$ given $\type$ and $\vec{t}_u$, where 
$\vec{t}_u$ denotes the vector containing types of nodes $v_1, \ldots, v_{\ell}$.
First, note that the node pairs $(j_1,j_2)$ given in (\ref{eq:to_compare_C_2}) are all in the set $T=\{\ell+1, \ldots, \ell+r\}$. Thus, the $\ii[E_{j_1j_2}]$ terms that appear in (\ref{eq:to_compare_C_2})
are disjoint from the  $\ii[E_{ij}]$ terms that appear in (\ref{eq:to_compare_C_3}). Second, the relations given in
 (\ref{eq:to_compare_C_2})
and
 (\ref{eq:to_compare_C_3})
constitute non-decreasing mappings of the rvs $\ii[E_{j_1j_2}]$ and
 $\ii[E_{ij}]$, respectively. 
 Thus, from the disjoint monotone aggregation property \cite[p. 35]{dubhashi2009concentration} of negative association, it  follows that the rvs $\ii[B]$ and $\ii[C]$ are conditionally negatively associated given $\type$ and $\vec{t}_u$ (equivalently, given the types of  nodes $v_1, \ldots, v_{\ell+r}$). Conditioning on $\vec{t}_u$, we  get
 \begin{align}
      &\pr[~B,C | \type]\nonumber \\
     &= \E\left[~\ii\left[B,C\right] ~~\big|~\type \right]
     \nonumber \\
     &=\E \left[~\E\left[ \ii [B] \ii[C] ~~~\big|~~ \type , \tu \right] ~\bigg|~ \type \right]\nonumber\\
     & \leq \E \left[~\E \left[ \ii[B] ~~~\big|~~ \type , \tu \right] \E \left[\ii [C] ~~~\big|~~ \type , \tu \right]
     ~~~\bigg|~ \type \right]
     \label{eq:intermediate_last_lemma}\\
     & = \E \left[~\pr \left[ C ~~\big|~~ \type \right]\E \left[~\ii[B] ~~\big|~~ \type , \tu \right] 
     ~~~\bigg|~ \type \right]
        \label{eq:intermediate2_last_lemma}\\
        & = \pr \left[ C ~~|~~ \type \right] \E \left[~\E \left[~\ii[B] ~~\big|~~ \type , \tu \right] 
     ~~~\bigg|~ \type \right]
     \nonumber \\
      & = \pr \left[ C ~~\big|~~ \type \right] \pr \left[ B ~~\big|~~ \type  \right] \nonumber
 \end{align}
where (\ref{eq:intermediate_last_lemma}) follows from conditional negative association of $\ii[B]$ and $\ii[C]$ given $\type$ and $\vec{t}_u$, and
(\ref{eq:intermediate2_last_lemma}) follows from the fact that event $C$ is independent from $\vec{t}_u$. 
\myendpf

\subsection{\bf Upper bounds for $\pr[\alr]$ }
\label{sec:subsec_ub_alr}
We have now obtained all the necessary bounds concerning the events $B$ and $C$ to obtain  upper bounds on $\alr$ tight enough to establish
(\ref{eq:kcon_suff_cond_new}); as already discussed, establishing (\ref{eq:kcon_suff_cond_new})  in turn completes the proof of the one-law for $k$-connectivity. In what follows, we will need to derive different upper bounds for $\alr$ in different ranges of $r = 2, \ldots, \lfloor \frac{n-\ell}{2} \rfloor$ to be considered in the sum appearing at (\ref{eq:kcon_suff_cond_new}).
Recall that $\alr= B \cap C \cap D = B \cap C \cap \dts \cap \dst$.
Among the events $B,C, \dts$ and $\dst$, the event $\dst$ depends exclusively on the choices made by nodes in $S$ while events $C$ and $\dts $ depend exclusively on choices made by nodes in $T$. However, the event $B$ depends on choices made by nodes in $T$ and $U$ and is thus dependent on the type of nodes in $T$ and $U$ with Lemmas~\ref{lem:kcon_bc_corr} and \ref{lem:kcon_b_corr} describing the  correlations. Our strategy for deriving an upper bound hinges on the selection of the subset of events which the yield the tightest bounds as $r$ varies. We partition the range of indices in the summation in (\ref{eq:kcon_suff_cond_new}) as outlined below.


\textbf{Range 1 (Small $r$):} When $2 \leq r \leq K_n-\ell$, for $S$ and $T$ to be isolated, all nodes in $T$ must be type-1, i.e., the occurrence of event $\tone$ is a necessary condition for the event $\alr$. If $\tone$ did not occur, then a type-2 node in $T$ could pick at most $r-1+\ell \leq K_n-1$ neighbors from among nodes in $T\cup U$ (distinct from itself). In that case, it will be forced to select at least one neighbor from $S$ contradicting the event $D$ (i.e., that there are no edges between $T$ and $S$).  
Noting that the probability of the event that all nodes in $T$ are type-1 is $\mu^r$, we have
\begin{align}
    \pr[B,C,D]&=\mu^r\pr[B,C,D~|~\tone] \nonumber\\
    &\leq \mu^r\pr[B,C,\dst~|~\tone] \label{eq:kcon_r1_1}\\
    &=\mu^r\pr[B,C~|~\tone] \pr[\dst]
    \label{eq:kcon_r1_2}\\
    &\leq \mu^r \pr[B~|~\tone] \pr[C~|~\tone]\pr[\dst]
    \label{eq:kcon_r1_3}
\end{align}
where (\ref{eq:kcon_r1_1}) follows from 
$\dst \subseteq{D}$,  (\ref{eq:kcon_r1_2}) follows from  independence of $\dst$ and $B,C,\tone$, and (\ref{eq:kcon_r1_3}) follows from Lemma~\ref{lem:kcon_bc_corr} with $t=1$. 



\textbf{Range 2 (Intermediate $r$):}
When $\K-\ell+1 \leq r \leq \lfloor \frac{n}{\K-1} \rfloor$, it is no longer needed to have $\tone$ for $\alr$ to take place. We instead condition on the increased likelihood of events $B$ and $C$ under the condition that all nodes in $T$ are of type-2 as follows.
\begin{align}
   \pr[B,C,D] &\leq \pr[B,C,\dst],\nonumber \\
   &=\pr[B,C] \pr[\dst],  \label{eq:intermediate_range_2} \\
      &\leq \pr[B,C~|~\ttwo] ~\pr[\dst],\label{eq:kcon_r2_1}\\
        &\leq \pr[B~|~\ttwo]~ \pr[C~|~\ttwo]~ \pr[\dst],\label{eq:kcon_r2_1_OY}
\end{align}
where (\ref{eq:intermediate_range_2}) follows from the independence of 
$\dst$ and $B,C$,
(\ref{eq:kcon_r2_1}) follows from Lemma \ref{lem:kcon_coupling_t2}, and (\ref{eq:kcon_r2_1_OY}) follows from  Lemma \ref{lem:kcon_bc_corr}.

\textbf{Range 3 (Large $r$):} When $\lfloor \frac{n}{\K-1} \rfloor +1 \leq r \leq \lfloor \frac{n-\ell}{2} \rfloor$, the number of nodes in $T$ is significantly large and the bound obtained in Lemma~\ref{lem:kcon_tree_bound} is no longer tight. 
Moreover, with the large number of nodes in $T$, the event $B$ that all nodes in B have a neighbor in $T$ becomes {\em highly} likely. Therefore, in this case, we consider both events $\dst$ and $\dts$ to get a tight upper bound for $B \cap C \cap D$. 
\begin{align}
    \pr[B,C,D] \leq \pr[D] = \pr[\dst, \dts] = \pr[\dst]  \pr[\dts]
\end{align}
since events $\dst$ and $\dts$ are independent.

Based on the preceding discussion, we partition the summation in (\ref{eq:kcon_suff_cond_new}) into three partial sums corresponding to Regimes 1,2 and 3 
In particular, we have
\begin{align}
   & \sum_{r=2}^{\floor{(n-\ell)/2}} {n\choose \ell}{n-\ell\choose r} \pr[B,C,D]    \label{eq:kcon_splitcases} \\
   & \quad \leq  \sum_{r=2}^{K_n-\ell} {n\choose \ell}{n-\ell\choose r} \mu^r \pr[B~|~\tone] \pr[C~|~\tone]\pr[\dst]\nonumber\\
   & \qquad   +\sum_{r=K_n-\ell+1}^{\floor{n/(\K-1)}} {n\choose \ell}{n-\ell\choose r} \pr[B~|~\ttwo] \pr[C~|~\ttwo] \pr[\dst]\nonumber\\
   & \qquad  +\sum_{\floor{n/(\K-1)}+1}^{\floor{(n-\ell)/2}} {n\choose \ell}{n-\ell\choose r} \pr[\dst] \pr[\dts]. 
   \nonumber 
\end{align}
The proof of (\ref{eq:kcon_suff_cond_new}) is completed in the Appendix A.4 by showing that these three partial sums approach zero as $n\rightarrow \infty$;
see Appendix A.4.\\

\section*{\textbf{ \large Appendix A.2: Proof of Lemma \ref{lem:coupling_summary}}}
\label{sec:proof_of_reduction_step}
\setcounter{subsection}{0}

Consider any scaling $\K: \N_0 \rightarrow \N_0$ such that the corresponding sequence $\g_n$ defined through (\ref{eq:kcon_scaling}) satisfies $\limit \g_n =\infty$. 
The next result shows that for any such scaling, an admissible scaling can be constructed with the corresponding parameters satisfying a useful bound. 
\begin{proposition}[Existence of Admissible Scaling]
Consider a scaling $K_n: \N_0 \rightarrow \N_0$ and a sequence $\g_n: \N_0 \rightarrow \R$ defined through (\ref{eq:kcon_scaling}) satisfying $\g_n \rightarrow +\infty$. Then, there exists an admissible scaling $\tilde{K}:\N_0 \rightarrow \N_0$ such that $\tilde{K}_n \leq \K$ for all $n \geq 2$, and the corresponding $\tilde{\g}_n$ defined through
\begin{align}
\tilde{K}_n=\frac{\log n +(k-2)\log \log n}{1-\mu}+\tilde{\g}_n,
\label{eq:kcon_scaling_tilde}
\end{align}
satisfies $\limit \tilde{\g}_n =\infty$. 
\label{prop:kcon_scalingexist}
\end{proposition}

\textbf{Proof of Proposition~\ref{prop:kcon_scalingexist}}
We prove this Proposition by constructing a scaling $\tilde{K}_n$ as follows. Let
\begin{align}
\tilde{K}_n:= \min\left \{ \left \lceil\frac{2 \log n}{1-\mu} \right \rceil, K_n\right \}, \quad n=2,3, \ldots  . \label{eq:kcon_ktilde}
\end{align}
By virtue of this definition, we have $\tilde{K}_n \leq {K_n}$
for all $n$.
Also, the mapping $\tilde{K}:\N_0 \rightarrow \N_0$ is a scaling with $2 \leq \tilde{K}_n < n \ \forall n \geq 2$.
From (\ref{eq:kcon_scaling_tilde}) and (\ref{eq:kcon_ktilde}), we have 
\begin{align}
\tilde{\g}_n= \min\left \{\lceil \log n - (k-2) \log \log n \rceil , \g_n\right \}. 
\label{eq:kcon_gammatilda_def}
\end{align}
Since $\limit \g_n = +\infty$, it is easy to see that 
$\limit \tilde{\g}_n =+\infty$. Also, since $\tilde{\g}_n \leq  \lceil \log n \rceil$ for all $n$, we have $\gamma_n = O(\log n)$.
Consequently, we see that the auxiliary scaling $\tilde{K}:\N_0 \rightarrow \N_0$ is indeed {\em admissible} as per Definition~\ref{def:kcon_admissible}. We note that the same parameter $\mu$ is used under both scalings.
\myendpf

{}
 
\vspace{2mm}
\noindent{\bf A reduction step.}
Let the inhomogeneous random K-out graph with $n$ nodes and parameters $\tilde{K}_n,\mu$ with $\tilde{K}_n$ defined through (\ref{eq:kcon_ktilde}) be denoted as ${\mathbb{H}(n;\mu,\tilde{K}_n)}$. Next, we present a way to infer the one-law for $k$-connectivity of $\hh$ from the connectivity of ${\mathbb{H}(n;\mu,\tilde{K}_n)}$ in the regime when $\gamma_n \rightarrow + \infty$ through the succeeding Proposition. 

\begin{proposition}[Coupling]
Consider a scaling $K: \N_0 \rightarrow \N_0$, then for any scaling $\widetilde{K}: \N_0 \rightarrow \N_0$ such that $\tilde{K}_n \leq K_n$ for all $n$, we have
\begin{align}
   & \pr[\hh \textrm{is $k$-connected} ]  \nonumber \\
   &\quad \geq \pr[{\mathbb{H}(n;\mu,\tilde{K}_n)} \textrm{is $k$-connected}].
   \nonumber
\end{align}
\label{prop:kcon_coupling}
\end{proposition}{}

\textbf{Proof of Proposition~\ref{prop:kcon_coupling}}
The proof involves showing the existence of a {\em coupling} between the graphs $\hh$ and $\hhc$ such that the edge set of $\hhc$ is contained in the edge set of $\hh$. The proof hinges on the observation that $k$-connectivity is a monotone-increasing property, i.e., a property which holds upon addition of edges to the graph (see \cite[p.~13]{coupling2017}). We will in fact show that 
\begin{align}
    &\pr[\hh \textrm{ has property } \mathcal{P} ]  \nonumber \\
   &\qquad \geq \pr[{\mathbb{H}(n;\mu,\tilde{K}_n)} \textrm{ has property } \mathcal{P}] \label{eq:monotone property coupling}
\end{align}
for any monotone property $\mathcal{P}$. 
In order to prove that the edge set of $\hhc$ is contained in $\hh$, we show that we can construct $\hh$ by adding edges to $\hhc$ as follows. Recall that during the construction of $\hh$, each node is first assigned a type corresponding to which it chooses neighbors uniformly at random. In particular, type-1 (resp., type-2) nodes pick 1 (resp., $\K$) nodes. An equivalent way to construct $\hh$ is as follows. The nodes are first initialized as type-1 (resp, type-2) independently with probability $\mu$ (resp., $1-\mu$). In the first round, type-1 (resp., type-2) nodes pick 1 (resp., $\tilde{K}_n$) neighbors. In the second round, each type-2 node picks additional $\K - \tilde{K}_n \geq 0$ neighbors chosen uniformly at random from the remaining $n-1-\tilde{K}_n$ nodes that it did not pick in the first round. The orientations of the edges drawn in the two rounds are ignored to yield $\hh$. From this construction, it is evident that the edge set of $\hhc$ is contained in the edge set of $\hh$.
Through this coupling argument, we see that (\ref{eq:monotone property coupling}) holds for any monotone increasing property $\mathcal{P}$. 
Since $k$-connectivity is monotonic-increasing upon addition of edges, the proof of Proposition \ref{prop:kcon_coupling} is completed.
\myendpf

Now that we have established Propositions~\ref{prop:kcon_scalingexist} and \ref{prop:kcon_coupling}, we can proceed with proving Lemma~\ref{lem:coupling_summary}.
[Proof of Lemma~\ref{lem:coupling_summary}]
Suppose, for any given parameters $(\mu, \K )$ the sequence $\g_n$ defined through (\ref{eq:kcon_scaling}) is such that $ \g_n \rightarrow  +\infty$. From Proposition~\ref{prop:kcon_scalingexist}, there exists an admissible scaling $\tilde{K}:\N_0 \rightarrow \N_0$ such that $\tilde{K}_n \leq \K, \forall n$ and the corresponding $\tilde{\gamma}_n \rightarrow + \infty$. If  the conditional statement in Lemma \ref{lem:coupling_summary} holds, i.e., if we have
\[
 \limit  \pr[\mathbb{H}(n;\mu, \tilde{K}_n) \textrm{ is $k$-connected } ] =1,
 \]
then it follows from Proposition~\ref{prop:kcon_coupling} that
\[
\limit  \pr[\hh \textrm{ is $k$-connected } ] =1.
\]
This completes the proof of Lemma~\ref{lem:coupling_summary}.
\myendpf

{}



\section*{\textbf{ \large Appendix A.3: Some useful facts}}
Here, we present some facts which will be frequently invoked in the succeeding analysis.
Consider an admissible scaling 
such that $\g_n$ defined through (\ref{eq:kcon_scaling}) satisfies $\g_n \rightarrow +\infty$. 
From the Definition~\ref{def:kcon_admissible} of admissible scaling, we  have $\g_n = \OO(\log n)$. Consequently, from (\ref{eq:kcon_scaling}) it is plain that we have
\begin{align}
K_n = \Theta (\log n).
\label{eq:kcon_knlogn}
\end{align}
under the assumptions enforced in Proposition~\ref{prop:kcon_thm1}.\\
For all $x \in \R$, we have
\begin{align}
    1  \pm x  &\leq e^{\pm x}. \label{eq:kcon_1pmx}
\end{align}
For $0 \leq x \leq 1$ we have
\begin{align}
    1 - \frac{x}{2} &\geq e^{-x}, \quad 0 \leq x \leq 1. \label{eq:kcon_xby2}
\end{align}
Moreover, for $0 \leq x < 1$ and for a sequence $y = 0,1,2\dots,$ we have
\begin{align}
  1-xy \leq  (1-x)^y \leq 1-xy+\frac{1}{2}x^2y^2.
   \label{eq:kcon_junfact}
\end{align}
A proof of this fact can be found
in \cite[Fact 2]{ZhaoYaganGligor}.
For $0\leq m \leq n_1 \leq n_2$,  $m, n_1,n_2 \in \N_0$, we have
For $0\leq m \leq n_1 \leq m, ~~  m,n_1,n_2 \in \N_0$,
\begin{align}
    \frac{{n_1 \choose m}}{{n_2 \choose m}}=\prod\limits_{i=0}^{m-1} \left( \frac{n_1-i}{n_2-i}\right) \leq \left( \frac{n_1}{n_2}\right)^m. \label{eq:kcon_choosefrac}
\end{align}
Using(\ref{eq:kcon_choosefrac}) and (\ref{eq:kcon_1pmx}), we have
\begin{align}
   \frac{{n-\ell \choose r}}{{n \choose r}} \leq \left(\frac{n-\ell}{n}\right)^r \leq \exp\left\{\frac{-r \ell}{n}\right\}.
   \label{eq:kcon_stdn-lchooser}
\end{align}
From \cite[Fact 4.1]{eletrebycdc2018}, we have that for $r=1,2,\dots, \lfloor \frac{n}{2} \rfloor$, we have
\begin{align}
  {n \choose r} \leq \left(\frac{n}{r} \right)^r  \left(\frac{n}{n-r} \right)^{n-r} 
  \label{eq:kcon_cdc18fact}
\end{align}
Combining (\ref{eq:kcon_stdn-lchooser}) and (\ref{eq:kcon_cdc18fact}), we get
\begin{align}
  {n-\ell \choose r} &\leq \left(\frac{n}{r} \right)^r  \left(\frac{n}{n-r} \right)^{n-r}  \hspace{-1mm} \exp\bigg \{\frac{-r \ell}{n}\bigg \} \nonumber \\
  &\leq \left(\frac{n}{r} \right)^r  \left(\frac{n}{n-r} \right)^{n-r}. \label{eq:kcon_n-lchooser}
\end{align}
For  any $\ell=0,1, \ldots$ we have
\begin{align}
    {n\choose \ell} = \frac{n^\ell}{\ell!} (1+\oo(1)). \label{eq:kcon_nchoosel}
\end{align}
Recall from Table~\ref{tab:event_notation} that $\mathcal{X}_t$ denotes the event that all nodes in $T$ are type-$t$ where $t=1,2$. Note that the event $\{\eij~|~ \text{$v_i, v_j \in T$ are type-$t$} \}$ is same as the event $\{\eij ~|~ \text{$v_i, v_j \in T$},\mathcal{X}_t\}$. Therefore,
\begin{align}
&\pr[\eij ~|~ \text{$v_i, v_j \in T$}, \mathcal{X}_t]\nonumber
\\
&= 1 - (1-\pr[i \in \Gamma_{n,j} | \text{$v_i, v_j \in T$}, \mathcal{X}_t]) \nonumber\\
& \qquad  \quad \cdot(1-\pr[j \in \Gamma_{n,i} | \text{$v_i, v_j \in T$}, \mathcal{X}_t])\nonumber \\
&=1-\left(1-\dfrac{K_t}{n-1}\right)^2
\nonumber \\
&=\dfrac{2K_t}{n-1}-\left(\dfrac{K_t}{n-1}\right)^2.
\label{eq:kcon_ijt}\end{align}
Moreover,
\begin{align}
&\pr[\eij~|~ \text{$v_j$ is type-$t$}]
\nonumber \\ 
&= 1 -\left( 1-\pr[i \in \Gamma_{n,j}~|~ \text{$v_j$ is type-$t$}] \right)  (1-\pr[j \in \Gamma_{n,i}])\nonumber \\
&=1-\left(1-\dfrac{K_t}{n-1}\right)\left(1-\dfrac{\kk}{n-1}\right) \nonumber\\
&=\dfrac{\kk}{n-1}+\dfrac{K_t}{n-1}-\dfrac{\kk K_t}{(n-1)^2}.
\label{eq:kcon_jt}\end{align}

\section*{\textbf{\large Appendix A.4: Proof of Proposition~\ref{prop:kcon_thm1}}}
\label{sec:kcon_mainProofproposition}
\setcounter{subsection}{0}

In this section we show that each of the partial sums corresponding to the three regimes outlined in Appendix A.1 (see  (\ref{eq:kcon_splitcases}))  approach zero as $n$ gets large. This then yields the one-law in Theorem~\ref{theorem:kcon} through the sufficient condition (\ref{eq:kcon_suff_cond_new}). Recall that we have 
(\ref{eq:kcon_knlogn}) under the assumptions enforced assumptions on the scaling $\K: \N_0 \rightarrow \N_0$ (i.e., that $\gamma_n=\omega(1)$, and $\gamma_n=O(\log n)$). 


\subsection{\bf Range 1: $2 \leq r \leq K_n-\ell$}
\label{subsec:regime1}
In this regime we evaluate the following partial sum in (\ref{eq:kcon_splitcases}) corresponding to $2 \leq r \leq K_n-\ell$. 
\begin{align}
    &\sum_{r=2}^{K_n-\ell} {n\choose \ell}{n-\ell\choose r} \mu^{r} \pr[B~|~\tone] \pr[C~|~\tone]\pr[\dst]. \label{eq:kcon_sumregime1}
\end{align}
Our strategy involves obtaining upper bounds on $\pr[B~|~\tone]$, $\pr[C~|~\tone]$ and $\pr[\dst]$.
We first upper bound $\pr[B~|~\tone]$ using Lemma~\ref{lem:kcon_b_corr} and (\ref{eq:kcon_jt}) with $t=1$ as follows.
\begin{align}
    &\pr[B~|~\tone] \nonumber \\
    &\leq r^\ell \left(\frac{\kk}{n-1}+\frac{1}{n-1}-\frac{\kk}{(n-1)^2}\right)^\ell, \nonumber \\
        &= r^\ell \left(\frac{\kk}{n}\right)^\ell \left(\frac{n}{n-1}\right)^{\ell} \left(1+\frac{1}{\kk}-\frac{1}{n-1}\right)^\ell, \nonumber \\
    &={\color{black}r^\ell \left(\frac{\kk}{n}\right)^\ell (1+\oo(1))} \label{eq:kcon_case1_pb} %
\end{align}
where (\ref{eq:kcon_case1_pb}) follows from (\ref{eq:kcon_knlogn})  and the fact that
\[
\left(\frac{n}{n-1}\right)^{\ell} \leq \exp\left \{\frac{\ell}{n-1} \right \} = 1+o(1)
\]
since $\ell$ is finite $(0 \leq \ell \leq k-1)$.

Next, we upper bound $\pr[C~|~\tone]$ using Lemma~\ref{lem:kcon_tree_bound} with $t=1$ and (\ref{eq:kcon_ijt}).
\begin{align}
 \pr[C~|~\tone]&\leq r^{r-2} \left(\frac{2}{n-1} - \frac{1}{(n-1)^2} \right)^{r-1}
 \nonumber\\
 &\leq r^{r-2} \left(\frac{2}{n-1} \right)^{r-1}
 \nonumber\\
 & \leq  r^{r-2} \left(\frac{2}{n} \right)^{r-1} \left(\frac{n}{n-1}\right)^{r-1}
 \nonumber\\
&\leq r^{r-2} \left(\frac{2}{n} \right)^{r-1} \exp\left\{\frac{r-1}{n-1}\right \}
\label{eq:kcon_case1_pc_osy}
\\
& = r^{r-2} \left(\frac{2}{n} \right)^{r-1} (1+o(1))
\label{eq:kcon_case1_pc}
\end{align}
where (\ref{eq:kcon_case1_pc_osy}) follows from (\ref{eq:kcon_1pmx}), and
(\ref{eq:kcon_case1_pc}) follows from the fact that $r \leq K_n =O(\log n)$ on the range considered here. 

Recall that $\dst$ is the event that nodes in $S$ do not choose a neighbor in $T$. For each $r=2,\dots, \lfloor \frac{n}{2} \rfloor$, we can condition on the types of nodes in $S$ to get
\begin{align}
 &\pr[\dst]\nonumber \\
 &=\left( \mu \left(1-\dfrac{r}{n-1}\right)+
 (1-\mu) \dfrac{{n-r-1 \choose \K}}{{n-1 \choose \K}} \right)^{n-(\ell+r)}  \label{eq:kcon_dstcommon_firststep}\\
  &\leq \left( \mu \left(1-\dfrac{r}{n-1}\right)+
 (1-\mu) \left(1-\dfrac{r}{n-1}\right)^{\K} \right)^{n-(\ell+r)} \label{eq:kcon_case1_11}\\
  &= \left(1-\dfrac{r}{n-1}\right)^{n-(\ell+r)} \nonumber\\
   & \quad \cdot\left( \mu +
 (1-\mu) \left(1-\dfrac{r}{n-1}\right)^{\K-1} \right)^{n-(\ell+r)}
 \nonumber\\
   &\leq \left(1-\dfrac{r}{n}\right)^{n-(\ell+r)} \left( \mu +
 (1-\mu) \left(1-\dfrac{r}{n}\right)^{\K-1} \right)^{n-(\ell+r)}
 \nonumber\\
   &= \left(1-\dfrac{r}{n}\right)^{n-(\ell+r)} \nonumber\\
   & \quad \cdot\left( 1- (1-\mu) \left(1-\left(1-\dfrac{r}{n}\right)^{\K-1}\right) \right)^{n-(\ell+r)}\label{eq:kcon_dstforregime3}
   \end{align}
where (\ref{eq:kcon_case1_11}) follows from (\ref{eq:kcon_choosefrac}). Note that for an admissible scaling, $\K= \OO{(\log n)}$. Thus, for $r \leq \K - \ell$, we have $\frac{r}{n}=\oo(1)$ and $\frac{r(K_n-1)}{n}=o(1)$. Using (\ref{eq:kcon_junfact}) with $x=\frac{r}{n}$ we get
\begin{align}
&\pr[\dst] \nonumber \\
 &\leq \bigg(1-\dfrac{r}{n}\bigg)^{n-(\ell+r)}  \cdot\bigg( 1- (1-\mu)\bigg(1-\bigg(1-\dfrac{r(\K-1)}{n} \nonumber\\
 & \quad +\dfrac{r^2 (\K-1)^2}{2n^2} \bigg)\bigg) \bigg)^{n-(\ell+r)} \nonumber \\
&=\left(1-\dfrac{r}{n}\right)^{n-r}  \left(1-\dfrac{r}{n}\right)^{-\ell}  \nonumber \\
        & \quad \cdot\left(1 - (1-\mu) \left(\dfrac{r (\K-1)}{n}-\dfrac{r^2 (\K-1)^2}{2n^2} \right) \right)^{n-(\ell+r)}
\nonumber  \\
&\leq (1+o(1))\bigg(1-\dfrac{r}{n}\bigg)^{n-r}\cdot\exp \bigg\{- (1-\mu)\bigg(n-(\ell+r)\bigg)\nonumber\\
        & \quad  \bigg(\dfrac{r (\K-1)}{n}-\dfrac{r^2 (\K-1)^2}{2n^2} \bigg) \bigg\}  
        \label{eq:kcon_common_ab_pd} \\
 &= (1+o(1))\bigg(1-\dfrac{r}{n}\bigg)^{n-r}  \cdot \exp \bigg\{ - (1-\mu)(n-(\ell+r))\nonumber \\& \quad \dfrac{r (\K-1)}{n}\bigg(1-\dfrac{r (\K-1)}{2n} \bigg) \bigg\} \label{eq:kcon_dstcommon} \\
&=  (1+o(1)) \left(1-\dfrac{r}{n}\right)^{n-r}  \nonumber \\
        & \quad \cdot\exp \left\{-(1-\mu)  r\left(\K-1 \right)  \left(1 - \frac{r (\K-1)}{2 n} \right) \right\}  \nonumber \\
        & \quad \cdot\exp \left\{(1-\mu)(\ell+r)  \dfrac{r (\K-1) }{n}   \left(1 - \frac{r (\K-1)}{2 n} \right) \right\} \nonumber,
\end{align}
where (\ref{eq:kcon_common_ab_pd}) follows from (\ref{eq:kcon_1pmx}) and the fact that $r \leq K_n =O(\log n)$. Further, when $\g_n \rightarrow \infty$, from scaling condition (\ref{eq:kcon_scaling}), we have $(1-\mu)(\K-1) \geq \log n$ for all $n$ sufficiently large. On that range, we have
\begin{align}
 & \pr[\dst]
 \nonumber \\
 &\leq (1+o(1)) \left(1-\dfrac{r}{n}\right)^{n-r}
 \exp \left\{ -r \log n \right\}\nonumber \\
 &\quad \cdot \exp  \left\{ \frac{r^2 \log n (\K-1)}{2 n}  \right\} \nonumber\\
 & \quad \cdot \exp \left\{(1-\mu) (\ell+r)\left(\dfrac{r (\K-1) }{n}\right)  \left(1 - \frac{r (\K-1)}{2 n} \right) \right\}
 \label{eq:kcon_case1_pd-1}
 \\
 & =(1+o(1))\left(\dfrac{n-r}{n}\right)^{n-r}
n^{-r}\left(1 + \oo{(1)} \right)
\label{eq:kcon_case1_pd}
 \end{align}
as we note that $\K= \OO{(\log n)}$ for an admissible scaling and $r \leq K_n$ on the range under consideration.
Combining (\ref{eq:kcon_n-lchooser}), (\ref{eq:kcon_nchoosel}), (\ref{eq:kcon_case1_pb}), (\ref{eq:kcon_case1_pc}), (\ref{eq:kcon_case1_pd}), we have for all $r =2, 3, \ldots, \K - \ell$ that
\begin{align}
&\mu^r {n\choose \ell}{n-\ell\choose r}  \pr[B~|~\tone] \pr[C~|~\tone]\pr[\dst] \nonumber\\
& \leq 
\mu^r 
\frac{n^\ell}{\ell!} 
\left(\frac{n}{r} \right)^r  \left(\frac{n}{n-r} \right)^{n-r}  
r^\ell \left(\frac{\kk}{n}\right)^\ell r^{r-2} \left(\frac{2}{n} \right)^{r-1}
\nonumber \\
& \quad \cdot 
\left(\dfrac{n-r}{n}\right)^{n-r} n^{-r}\left(1 + \oo{(1)} \right)\nonumber\\
& =
\frac{(2\mu)^r}{2 \ell !}
   \frac{ r^{\ell-2} \kk^\ell}{n^{r-1}} \left(1 + \oo{(1)} \right)
   \nonumber 
   \\
&=\frac{2 \mu^2}{ \ell !} \left(\frac{2\mu}{n}  \right)^{r-2}  \frac{r^{\ell-2} \kk^\ell }{n} \left(1 + \oo{(1)} \right)
\nonumber \\
& = \left(\frac{2\mu}{n}  \right)^{r-2}  \oo(1), \label{eq:kcon_case1_termbound}
\end{align}
where 
(\ref{eq:kcon_case1_termbound}) follows from $r \leq \K =O(\log n)$. In order to show that the summation (\ref{eq:kcon_sumregime1}) is $\oo(1)$, we upper bound it by an infinite geometric progression wherein each term of the geometric progression is non-negative and strictly less than 1. Note that since (\ref{eq:kcon_case1_termbound}) holds for all $r$ such that $2 \leq r \leq \K-\ell$, substituting in (\ref{eq:kcon_sumregime1}) we obtain
\begin{align}
    &\sum_{r=2}^{\K-\ell} \mu^r {n\choose \ell}{n-\ell\choose r} \pr[B~|~\tone] \pr[C~|~\tone]\pr[\dst]
    \nonumber \\
    & \leq \oo(1) \sum_{r=2}^{\K-\ell}\left(\frac{2\mu}{n}  \right)^{r-2}  
    \nonumber \\
    & \leq \oo(1) \sum_{r=0}^{\infty}\left(\frac{2\mu}{n}  \right)^{r}
    \nonumber \\
    & = \frac{1}{1-\frac{2 \mu}{n}}\oo(1) 
    \nonumber \\
    & =  \oo(1). \label{eq:kcon_sumregime1_0proved}
\end{align}
 
\subsection{\bf Range 2: $  \K-\ell < r \leq \lfloor{n/(\K-1)}\rfloor$}
\label{subsec:regime2}
Here, we consider the partial sum in (\ref{eq:kcon_splitcases}) corresponding to the range $\K-\ell < r \leq \lfloor{n/(\K-1)\rfloor}$, i.e., 
\begin{align}
\sum_{r=K_n-\ell+1}^{\floor{n/(\K-1)}} {n\choose \ell}{n-\ell\choose r} \pr[B~|~\ttwo] \pr[C~|~\ttwo] \pr[\dst] \label{eq:kcon_sumregime2}
\end{align}
From (\ref{eq:kcon_knlogn}), we know that $\K=\Theta (\log n)$ and therefore in this range 
$r \leq \lfloor{n/(\K-1)\rfloor} \leq \lfloor{n/2\rfloor}$. As noted in Appendix A.1, our strategy involves combining upper bounds on $\pr[B~|~\ttwo]$, $\pr[C~|~\ttwo]$ and $\pr[\dst]$ obtained using Lemmas~\ref{lem:kcon_b_corr}, \ref{lem:kcon_tree_bound} and \ref{lem:kcon_bc_corr}.
From Lemma~\ref{lem:kcon_b_corr} and (\ref{eq:kcon_jt}) with $t=2$,
\begin{align}
    \pr[B~|~\ttwo]
    &\leq  r^\ell \left(\frac{\kk}{n-1}+\frac{\K}{n-1}-\frac{\kk \K}{(n-1)^2}\right)^\ell
  \nonumber \\ 
    &\leq 
    r^\ell \left(\frac{\kk}{n-1}+\frac{\K}{n-1}\right)^\ell
    \nonumber
    \end{align}
    Substituting $\kk = \mu +(1-\mu)\K$, we get
    \begin{align}
    \pr[B~|~\ttwo]
    &\leq r^\ell  \left(\frac{(2-\mu)\K}{n-1}\right)^\ell \nonumber \\ 
    &\leq \left(1+\frac{\mu}{(2-\mu)\K}\right)^\ell
    \nonumber\\
    &=r^\ell \left(\frac{(2-\mu)\K}{n}\right)^\ell (1+\oo(1))
    \label{eq:kcon_regime2_b}
\end{align}
where (\ref{eq:kcon_regime2_b}) follows from (\ref{eq:kcon_knlogn}) and 
the fact that
$\left(n/(n-1)\right)^{\ell} =1+o(1)$.

Next, we bound $\pr[C~|~\ttwo]$ using Lemma~\ref{lem:kcon_tree_bound} and (\ref{eq:kcon_ijt}) with $t=2$. We get
\begin{align}
 \pr[C~|~\ttwo]&\leq r^{r-2} \left(\frac{2 \K}{n-1} - \frac{\K^2}{(n-1)^2} \right)^{r-1}
 \nonumber\\
 & = r^{r-2} \left(\frac{2 \K}{n} \right)^{r-1} \left(1+\frac{1}{n-1} -\frac{n \K }{2 (n-1)^2}\right)^{r-1}
\nonumber\\
& \leq r^{r-2} \left(\frac{2 \K}{n} \right)^{r-1} 
\label{eq:kcon_regime2_c}
\end{align}
where (\ref{eq:kcon_regime2_c}) follows from 
(\ref{eq:kcon_knlogn}).

Next, we find an upper bound on $\pr[\dst]$. 
Note that 
$\frac{r}{n} \leq 1/(\K-1) = o(1)$ in view of (\ref{eq:kcon_knlogn}).
Consequently, we can use (\ref{eq:kcon_junfact}) with $x=\frac{r}{n}$. Also, it still holds that
$(1-r/n)^{-\ell} = 1+o(1)$.
Thus, proceeding as in Range 1, we can upper bound $\pr[\dst]$ by undergoing the sequence of steps from (\ref{eq:kcon_dstcommon_firststep}) through (\ref{eq:kcon_dstcommon}) to obtain
\begin{align}
 & \pr[\dst] \nonumber\\
  &\leq (1+o(1)) \bigg(1-\dfrac{r}{n}\bigg)^{n-r}  \label{eq:kconregime2temp1}
\cdot \exp \bigg\{ - (1-\mu)(n-(\ell+r))  \nonumber\\
  & \quad  \dfrac{r (\K-1)}{n}\bigg(1-\dfrac{r (\K-1)}{2n} \bigg) \bigg\}. 
\end{align}
In this range, we have $r \leq \frac{n}{(\K-1)}$ and thus $\frac{r (\K -1 )}{2n} \leq \frac{1}{2}$, or equivalently $1-\frac{r (\K -1 )}{2n} \geq \frac{1}{2}$. Further, since $r \leq \frac{n-\ell}{2}$, we have $n-(\ell+r) \geq \frac{n-\ell}{2}$. Using these observations in (\ref{eq:kconregime2temp1}), we get
\begin{align}
 & \pr[\dst]
 \nonumber \\
  &\leq (1+o(1)) \left(1-\dfrac{r}{n}\right)^{n-r}  \nonumber\\
  & \quad \cdot \exp \left\{ -  \dfrac{(1-\mu)(n-\ell)r (\K-1)}{4n} \right\}
  \nonumber\\
  &= (1+o(1)) \left(1-\dfrac{r}{n}\right)^{n-r} 
  \exp \left\{ -  \dfrac{(1-\mu) r (\K-1)}{4} \right\}  
 \nonumber  \\ 
 & \qquad \cdot \exp \left\{  \dfrac{\ell (1-\mu) r (\K-1)}{4n} \right\}
  \nonumber 
  \\
 &= O(1) \left(1-\dfrac{r}{n}\right)^{n-r} 
  \exp \left\{ -  \dfrac{(1-\mu) r (\K-1)}{4} \right\},
  \label{eq:kcon_regime2_d}
\end{align}
where in the last step we used the fact that $\exp \left\{  \frac{\ell (1-\mu) r (\K-1)}{4n}\right\} =O(1)$ since $\frac{r (\K -1 )}{n} \leq 1$. 
Combining  (\ref{eq:kcon_n-lchooser}),
(\ref{eq:kcon_nchoosel}),
(\ref{eq:kcon_regime2_b}), (\ref{eq:kcon_regime2_c}) and (\ref{eq:kcon_regime2_d}), we get 
\begin{align}
&{n\choose \ell}{n-\ell\choose r}  \pr[B~|~\ttwo] \pr[C~|~\ttwo] \pr[\dst] \nonumber\\
& \leq 
O(1) \frac{n^\ell}{\ell!} 
\left(\frac{n}{r} \right)^r  \left(1-\frac{r}{n} \right)^{-(n-r)}  
r^\ell \left(\frac{(2-\mu)\K}{n}\right)^\ell r^{r-2} 
\nonumber\\
& \quad \cdot \left(\frac{2 \K}{n} \right)^{r-1} 
\left(1-\dfrac{r}{n}\right)^{n-r} 
  \exp \left\{ -  \dfrac{(1-\mu)r (\K-1)}{4} \right\} \nonumber\\
&= O(1) n r^{\ell-2} \K^{r+\ell-1}  2^{r}  \exp \left\{ -  \dfrac{(1-\mu) r (\K-1)}{4} \right\}. \label{eq:kcon_regime2_termbound}
\end{align}

Substituting (\ref{eq:kcon_regime2_termbound}) in (\ref{eq:kcon_sumregime2}), we get
\begin{align}
    &\sum_{r=K_n-\ell+1}^{\floor{n/(\K-1)}} {n\choose \ell}{n-\ell\choose r} \pr[B~|~\ttwo] \pr[C~|~\ttwo] \pr[\dst], \nonumber\\
    & =  O(1) n \K^{\ell-1} \nonumber\\ & \quad \cdot\sum_{r=K_n-\ell+1}^{\floor{n/(\K-1)}} r^{\ell-2} {2^{r} \K^{r}   \exp {\left\{ -  \dfrac{(1-\mu) r (\K-1)}{4} \right\}}. }
\nonumber \\ 
    &\leq  O(1) n \K^{\ell-1} \left(\frac{n}{\K-1}\right)^{\ell-2} 
    \nonumber \\
   & \quad 
   \cdot  \sum_{r=K_n-\ell+1}^{\floor{n/(\K-1)}}  
          {\left(2 \K   \exp \left \{ -  \dfrac{(1-\mu) (\K-1)}{4} \right \}\right)^r }
          \nonumber\\
     &= O(1) n^{\ell-1} \K  \nonumber \\ & \quad \cdot \sum_{r=K_n-\ell+1}^{\floor{n/(\K-1)}} 
          {\left(2 \K   \exp {\left\{ -  \dfrac{(1-\mu) (\K-1)}{4} \right\}}\right)^r }
          \nonumber \\
          & \leq  O(1) n^{\ell-1} \K \nonumber \\
          & \quad \cdot \sum_{r=K_n-\ell+1}^{\infty} 
          {\left(2 \K   \exp {\left\{ -  \dfrac{(1-\mu) (\K-1)}{4} \right\}}\right)^r }
         \label{eq:kcon_regime2_osy}
        \end{align} 
Since $\limit K_n  = \infty$ we have 
\begin{align}
   2 \K   \exp {\left\{ -  \frac{(1-\mu) (\K-1)}{4} \right\}} = \oo(1)
   \label{eq:kcon_regime2_gpterm}
\end{align}
so that the infinite sum appearing at (\ref{eq:kcon_regime2_osy}) is summable. 
   Thus, we get
    \begin{align}
     &\sum_{r=K_n-\ell+1}^{\floor{n/(\K-1)}} {n\choose \ell}{n-\ell\choose r} \pr[B~|~\ttwo] \pr[C~|~\ttwo] \pr[\dst],\nonumber\\
    &=O(1) n^{\ell-1} \K  \quad 
    {\left(2 \K   \exp {\left\{ -  \dfrac{ (1-\mu)(\K-1)}{4} \right\}}\right)^{K_n-\ell+1} }
    \nonumber\\
  & = O(1) 
   \exp \Bigg\{(\ell-1) \log n + \log \K  + (\K-\ell+1)\log 2\K
 \nonumber \\&\qquad \qquad \ \ \ \ -\frac{(1-\mu)(\K-1)(\K-\ell+1)}{4} \Bigg\}
 \nonumber\\
   &= O(1) \exp \left \{- \frac{\K^2  (1-\mu)}{4} (1+ \oo(1))\right\}
     \label{eq:kcon_sumregime2_0proved}
      \\ 
   & = \oo(1)
   \label{eq:kcon_sumregime2_0proved_osy}
\end{align}
where  (\ref{eq:kcon_sumregime2_0proved}) and (\ref{eq:kcon_sumregime2_0proved_osy}) both follow from $\K =\Theta(\log n)$, respectively. 

\subsection{\bf Range 3: $\lfloor{n/(\K-1)}\rfloor+1 < r \leq \floor{(n-\ell)/2}$}
\label{subsec:regime3}
Here, we will consider the
partial sum in  
(\ref{eq:kcon_splitcases}) with index $r$ over the range $\lfloor{n/(\K-1)}\rfloor+1 < r \leq \floor{(n-\ell)/2}$; i.e.,
the term
\begin{align}
   &\sum_{r=\lfloor{n/(\K-1)}\rfloor+1}^{\floor{(n-\ell)/2}} {n\choose \ell}{n-\ell\choose r} \pr[\dst] \pr[\dts]. \label{eq:kcon_regime3_main}
\end{align}
Note that since $\K=\Theta(\log n)$, this range is non-empty. Conditioning on the types of nodes in $T$, it is easy to verify that for each $r$ in $2,\dots, \lfloor \frac{n}{2} \rfloor$ we have
\begin{align}
    &\pr[\dts]  \nonumber\\
    &=
 \left( \mu \dfrac{{r-1+\ell \choose 1}}{{n-1 \choose 1}}+
 (1-\mu) \dfrac{{r-1+\ell \choose \K}}{{n-1 \choose \K}} \right)^{r}
 \nonumber\\
 &\leq
 \left( \mu \left( \dfrac{{r-1+\ell }}{{n-1 }} \right)+
 (1-\mu) \left(\dfrac{{r-1+\ell }}{{n-1 }} \right)^{\K} \right)^{r}
 \label{eq:kcon_regime3_t1}\\
& =\mu^r   \left(\dfrac{r-1+\ell }{n-1 } \right)^{r}
 \left( 1+ \frac{1-\mu}{\mu} \left(\dfrac{{r-1+\ell }}{{n-1 }} \right)^{\K-1} \right)^{r}
 \nonumber\\
 & \leq \mu^r   \left(\dfrac{r+\ell }{n } \right)^{r}
 \left( 1+ \frac{1-\mu}{\mu} \left(\dfrac{{r+\ell }}{{n }} \right)^{\K-1} \right)^{r} \label{eq:kcon_regime3_t2}\\
 & \leq \mu^r   \left(\dfrac{r+\ell }{n } \right)^{r}
\exp{ \left\{ \frac{1-\mu}{\mu} r \left(\dfrac{{r+\ell }}{{n }} \right)^{\K-1} \right\}
} \label{eq:kcon_case3_pdts}
\end{align}
where (\ref{eq:kcon_regime3_t1})  and (\ref{eq:kcon_case3_pdts}) follow from (\ref{eq:kcon_choosefrac}) and (\ref{eq:kcon_1pmx}), respectively, and (\ref{eq:kcon_regime3_t2}) follows from the  $r+\ell \leq n$.

We  bound $\pr[\dst]$ by using the sequence of steps from (\ref{eq:kcon_dstcommon_firststep}) through
(\ref{eq:kcon_dstforregime3}) as in range 1. We get
\begin{align}
    &\pr[\dst]
    \nonumber\\
   & \leq \left(1-\dfrac{r}{n}\right)^{n-(\ell+r)} \nonumber \\
   & \quad \cdot \left( 1- (1-\mu) \left(1-\left(1-\dfrac{r}{n}\right)^{\K-1}\right) \right)^{n-(\ell+r)}
       \nonumber\\
    &\leq \left(1-\dfrac{r}{n}\right)^{n-(\ell+r)} 
    \nonumber \\
   & \quad \cdot
   \left( 1- (1-\mu) \left(1-\exp\left\{\dfrac{-r({\K-1})}{n}\right\}\right) \right)^{n-(\ell+r)}
    \nonumber\\
     &\leq  \bigg(1-\dfrac{r}{n}\bigg)^{n-(\ell+r)}  \exp \bigg\{- (1-\mu){(n-(\ell+r))}\nonumber\\ &\quad \cdot \bigg(1-\exp\bigg\{\dfrac{-r({\K-1})}{n}\bigg\}\bigg)\bigg\} \label{eq:kcon_case3_pdst}
\end{align}
using (\ref{eq:kcon_1pmx}). 

Let $Z_{n,r}$ be defined as 
\begin{align}
   & Z_{n,r}= \nonumber\\
   &\exp{ \bigg\{ \frac{1-\mu}{\mu} r \bigg(\dfrac{{r+\ell }}{{n }} \bigg)^{\K-1}\bigg\}}  
   \quad \cdot\exp \bigg\{
- (1-\mu)\nonumber\\
&\quad{(n-(\ell+r))}\bigg(1-\exp\bigg\{\dfrac{-r({\K-1})}{n}\bigg\}\bigg)\bigg\}. \label{kcon_regime3_znr}
\end{align}

Substituting 
(\ref{eq:kcon_case3_pdts}), (\ref{eq:kcon_case3_pdst}), 
 and (\ref{kcon_regime3_znr}) in (\ref{eq:kcon_regime3_main}), we get
\begin{align}
   &\sum_{r=\frac{n}{\lfloor \K - 1}\rfloor+1}^{\floor{(n-\ell)/2}} {n\choose \ell}{n-\ell\choose r} \pr[\dst] \pr[\dts]
   \nonumber\\
 & \leq \sum_{r=\lfloor{n/(\K-1)}\rfloor+1}^{\floor{(n-\ell)/2}} {n\choose \ell}{n-\ell\choose r}
 \mu^r   \left(\dfrac{r+\ell }{n } \right)^{r} \nonumber\\ 
 & \quad \cdot \left(1-\dfrac{r}{n}\right)^{n-(\ell+r)}   Z_{n,r}
 \nonumber\\
 &\leq \sum_{r=\lfloor{n/(\K-1)}\rfloor+1}^{\floor{(n-\ell)/2}}\frac{n^\ell}{\ell!} (1+\oo(1))\left(\frac{n}{r} \right)^r  \left(\frac{n}{n-r} \right)^{n-r}  \mu^r  
 \nonumber \\ 
 &\quad \quad  \cdot    \left(\dfrac{r+\ell }{n } \right)^{r} \left(1-\dfrac{r}{n}\right)^{n-(\ell+r)}   Z_{n,r}
 \label{eq:kcon_regime3_sum_t2}\\
 &= \sum_{r=\lfloor{n/(\K-1)}\rfloor+1}^{\floor{(n-\ell)/2}} O(1) n^\ell \mu^r   \left(1+\frac{\ell}{r} \right)^r \left(1-\dfrac{r}{n}\right)^{-\ell}
 Z_{n,r}
 \nonumber\\ 
 & \leq \sum_{r=\lfloor{n/(\K-1)}\rfloor+1}^{\floor{(n-\ell)/2}} O(1)  n^\ell \mu^r 
 2^\ell e^\ell 
 Z_{n,r} \label{eq:kcon_regime3_sum_t3}\\
 &=\sum_{r=\lfloor{n/(\K-1)}\rfloor+1}^{\floor{(n-\ell)/2}} O(1) \mu^r n^\ell Z_{n,r}  \label{eq:kcon_regime3_sum_t4}
\end{align}
where (\ref{eq:kcon_regime3_sum_t2}) follows from 
(\ref{eq:kcon_n-lchooser}) and
(\ref{eq:kcon_nchoosel}),
 (\ref{eq:kcon_regime3_sum_t3}) is apparent from $r\leq n/2$ implying $(1-r/n)^{-\ell}\leq 2^\ell$ and  (\ref{eq:kcon_1pmx}) implying  $\left(1+\ell/r \right) \leq e^\ell$. 
 
 Next, we derive an upper bound for $ Z_{n,r}$.  
 Our goal is to show that $ Z_{n,r}$ goes to zero as $n\rightarrow \infty$ for each $r$ in $\lfloor{n/(\K-1)}\rfloor+1 < r \leq \floor{(n-\ell)/2}$. 
The approach used in this part is reminiscent of some of the techniques used for proving 1-connectivity of $\hh$ in \cite{eletrebycdc2018}. Recall that  $\ell \leq k$ where $k$ is finite, and $r/n \leq 1/2$ on the range considered. Further, we have 
$r+\ell \leq {(n+\ell)}/{2} \leq {(n+n/2)}/{2} = 3n/4$. Thus,
\begin{align}
r \left(\dfrac{{r+\ell }}{{n }} \right)^{\K-1} \leq \frac{n}{2} (0.75)^{\K-1}. \label{eq:kcon_case3_anr_1}
\end{align}
Recall that in Range 3, $\frac{r(\K-1)}{n}>1$ and therefore
\begin{align}
\exp \left\{\frac{-r(\K-1)}{n}\right\}<e^{-1}. \label{eq:kcon_case3_anr_2}
\end{align}
Using (\ref{eq:kcon_case3_anr_1}) and (\ref{eq:kcon_case3_anr_2}) in (\ref{kcon_regime3_znr}), we see that
\begin{align}
    Z_{n,r} & \leq \exp \bigg\{ \frac{(1-\mu)}{\mu} \frac{n}{2} \bigg(0.75 \bigg)^{\K-1}\nonumber\\
&- (1-\mu) {(n-(\ell+r))}\bigg(1-e^{-1}\bigg)\bigg\}
\end{align}

Further, noting that $n-(\ell+r) > (n-\ell)/2$, we get
\begin{align}
  &  Z_{n,r}
  \nonumber \\
& \leq \exp{ \left\{ \frac{(1-\mu)}{\mu} \frac{n}{2}\left(0.75 \right)^{\K-1}
- (1-\mu) \frac{n-\ell}{2}\left(1-e^{-1}\right)\right\}}
\nonumber\\
 & =\exp{ \left\{ -(1-\mu)\frac{n}{2}\left(1-e^{-1}- \frac{0.75^{\K-1}}{\mu} \right)\right\}} \nonumber\\ 
 & \quad \cdot \exp \left\{  \frac{(1-\mu) \ell (1-e^{-1})}{2} \right\} 
 \nonumber \\ 
& = \exp{ \left\{ -(1-\mu)\frac{n}{2}\left(1-e^{-1}- o(1) \right)\right\}}  O(1)
  \label{eq:kcon_regime3_t5_osy}
  \\ 
 & = \oo(1)
 \label{eq:kcon_regime3_t5}
\end{align}
where (\ref{eq:kcon_regime3_t5_osy}) is a consequence of $\K = O(\log n)$  and the fact that $\ell$ is finite. As before, we use an infinite geometric progression to upper bound the summation in (\ref{eq:kcon_regime3_sum_t4}) using 
(\ref{eq:kcon_regime3_t5}). Combining (\ref{eq:kcon_regime3_sum_t4}) and (\ref{eq:kcon_regime3_t5}), we obtain
\begin{align}
   & \sum_{r=\lfloor{n/(\K-1)}\rfloor+1}^{\floor{(n-\ell)/2}} {n\choose \ell}{n-\ell\choose r} \pr[\dst, \dts] 
   \nonumber\\
   & \qquad \leq o(1) \ n^\ell \sum_{r=\lfloor{n/(\K-1)}\rfloor+1}^{\infty} \mu^r  
  \nonumber\\
 \nonumber\\
        & \qquad  =  o(1) \  n^\ell     \mu^{n/(\K-1)}
        \nonumber\\
        & \qquad  =  o(1),
        \label{eq:kcon_sumregime3_0proved-1}
\end{align}
where (\ref{eq:kcon_sumregime3_0proved-1}) follows from the fact that
\[
n^\ell  \mu^{n/(\K-1)} =  n^\ell     \mu^{n/\Theta(\log n)} = o(1).
\]
in view of (\ref{eq:kcon_knlogn}), $\mu<1$, and $\ell$ being finite. 
We thus conclude that 
\begin{align}
    \sum_{r=\lfloor{n/(\K-1)}\rfloor+1}^{\floor{(n-\ell)/2}} {n\choose \ell}{n-\ell\choose r} \pr[\dst] \pr[\dts]
    = \oo{(1)}. \label{eq:kcon_regime3_final}
\end{align}

\textbf{Proof of Proposition~\ref{prop:kcon_thm1}}
Combining (\ref{eq:kcon_sumregime1_0proved}), (\ref{eq:kcon_sumregime2_0proved_osy}), and (\ref{eq:kcon_regime3_final}), it is evident that all of the three partial sums approach zero as $n$ approaches $\infty$ and thus we have proved the sufficient condition (\ref{eq:kcon_suff_cond_new}) for $k$-connectivity. The proof of Proposition~\ref{prop:kcon_constrained} is now complete.

\myendpf

\setcounter{subsection}{0}
\section*{\bf \Large Appendix B (Size of Giant Component)}
\vspace{15 pt}
In this section, we provide supplementary details for the proof of Theorem~\ref{theorem:gc}.
\subsection{\bf Proof of Proposition~\ref{prop:gcproofk3}}

Recall that we have
\begin{align}
\mathcal{Z}(x_n;\mu,K_n) & = \bigcap_{S \in \mathcal{P}_n: ~x_n\leq  |S| \leq \lfloor \frac{n}{2} \rfloor}  \left(\mathcal{E}_n({\mu},{K}_n; S)\right)\comp,
\nonumber 
\end{align}
where $\mathcal{Z}(x_n;\mu,K_n)$ denote the event that $\hh$ has no cut $S \subset \nodes$ with size  $x_n \leq |S| \leq n-x_n$. Taking the complement of both sides and using a union bound we get
\begin{align}
\pr\left[\left(\mathcal{Z}(x_n;\mu,K_n)\right)\comp\right] &\leq  \sum_{ S \in \mathcal{P}_n: x_n \leq |S| \leq \lfloor \frac{n}{2} \rfloor } \pr[ \mathcal{E}_n ({\mu},{K}_n; S) ] \nonumber \\
&=\sum_{r=x_n}^{ \left\lfloor \frac{n}{2} \right\rfloor }
\left ( \sum_{S \in \mathcal{P}_{n,r} } \pr[\mathcal{E}_n ({\mu},{K}_n; S)] \right ) \label{eq:BasicIdea+UnionBound},
\end{align}
where  $\mathcal{P}_{n,r} $ denotes the collection of all subsets of $\nodes$ with exactly $r$ elements.
For each $r=1, \ldots , n$, we simplify the notation by writing $\mathcal{E}_{n,r} ({\mu},{K}_n;S)=\mathcal{E}_n ({\mu},{K}_n ; \{ 1, \ldots , r \} )$. From the exchangeability of the node labels and associated random variables, we get
\[
\pr[ \mathcal{E}_n({\mu},{K}_n ; S) ] = \pr[ \mathcal{E}_{n,r}(\mu,\K) ], \quad S \in
\mathcal{P}_{n,r}.
\]
Noting that $|\mathcal{P}_{n,r} | = {n \choose r}$, we obtain
\begin{equation}
\sum_{S \in \mathcal{P}_{n,r} } \pr[\mathcal{E}_n ({\mu},{K}_n ; S) ] 
= {n \choose r} ~ \pr[\mathcal{E}_{n,r} ({\mu},{K}_n)].  \nonumber
\label{eq:ForEach=r}
\end{equation}
Substituting into (\ref{eq:BasicIdea+UnionBound}) we obtain 
\begin{align}
\pr\left[\left(\mathcal{Z}(x_n;\mu,K_n)\right)\comp\right] \leq \sum_{r=x_n}^{ \left\lfloor \frac{n}{2} \right\rfloor }
{n \choose r} ~ \pr[ \mathcal{E}_{n,r}({\mu},{K}_n) ] .
\label{eq:gc_zbound}
\end{align}

In view of (\ref{eq:gc_zbound}), the next proposition provides an upper bound on $\pr\left[\left(\mathcal{Z}(M;\mu,K_n)\right)\comp\right]$, 
i.e, the probability that there exists a cut with size in the range $[M,  n-M]$ for $\hh$.  Recall that $K_n = \omega(1)$ imparts 1-connectivity \cite{Rashad/Inhomo} whp and we focus on cases where $\K$ is a {\em bounded} sequence.


In view of (\ref{eq:gc_zbound}), the proof for Proposition~\ref{prop:gcproofk3} will follow upon showing
\begin{align}
 & \sum_{r=M}^{\floor{n/2}} {n \choose r}\pr[\sdcon]\nonumber \\
 &\leq \frac{\exp\{-M\left(\kk-1\right)(1-\oo(1))\}}{1-{\exp\{-\left(\kk-1\right)(1-\oo(1))\}}} + \oo(1) . \label{eq:propeqn}
 \end{align}{}
 
%


We have
 \begin{align}
 & {n \choose r}\pr[\sdcon]\nonumber \\
 &={n \choose r} \left( \mu \left(\dfrac{n-r-1}{n-1}\right)+(1-\mu) \dfrac{{n-r-1 \choose \K}}{{n-1 \choose \K}} \right)^{n-r}  \nonumber \\
  & \quad \cdot \left( \mu \left(\dfrac{r-1}{n-1}\right)+(1-\mu) \dfrac{{r-1 \choose \K}}{{n-1 \choose \K}} \right)^{r} \nonumber  \\
 &\leq{n \choose r}\left( \mu \left(1-\dfrac{r}{n-1}\right)+
 (1-\mu) \left(1-\dfrac{r}{n-1}\right)^{\K} \right)^{n-r}\nonumber \\
  &  \quad \cdot \left( \mu \left(\dfrac{r-1}{n-1}\right)+(1-\mu) \left(\dfrac{{r-1}}{{n-1 }} \right)^{\K} \right)^{r} \label{eq:gc_eq1} \\
   &\leq{n \choose r}\left( \mu \left(1-\dfrac{r}{n}\right)+
 (1-\mu) \left(1-\dfrac{r}{n}\right)^{\K} \right)^{n-r}\nonumber \\
  &  \quad \cdot \left( \mu \left(\dfrac{r}{n}\right)+(1-\mu) \left(\dfrac{{r}}{{n }} \right)^{\K} \right)^{r}  \nonumber  \\
  & \leq  \left(\frac{n}{r} \right)^r  \left(\frac{n}{n-r} \right)^{n-r}\left(1-\dfrac{r}{n}\right)^{n-r} \left( \dfrac{r}{n}\right)^r \nonumber \\
 & \quad \cdot \left( \mu+
 (1-\mu) \left(1-\dfrac{r}{n}\right)^{\K-1} \right)^{n-r}\nonumber \\
  & \quad \cdot \left( \mu +(1-\mu) \left(\dfrac{{r}}{{n }} \right)^{\K-1} \right)^{r} \label{eq:gc_eq2} \\
    & = \left( \mu+
 (1-\mu) \left(1-\dfrac{r}{n}\right)^{\K-1} \right)^{n}\nonumber \\
  & \quad \cdot \left(  \dfrac{\mu +(1-\mu) \left(\dfrac{{r}}{{n }} \right)^{\K-1}}{ \left( \mu+
 (1-\mu) \left(1-\dfrac{r}{n}\right)^{\K-1} \right)} \right)^{r} \nonumber \\
     & \leq  \left( \mu+
 (1-\mu) \left(1-\dfrac{r}{n}\right)^{\K-1} \right)^{n} \label{eq:gc_beforesplit}
    \end{align}
 where (\ref{eq:gc_eq1}) uses (\ref{eq:kcon_choosefrac}), (\ref{eq:gc_eq2}) follows from (\ref{eq:kcon_cdc18fact}) and (\ref{eq:gc_beforesplit}) is plain from the observation that $ r/n \leq 1/2$.
   \par We divide the summation in (\ref{eq:propeqn}) into two parts depending on whether $r$ exceeds ${n}/{\log n}$. The steps outlined below can be used to upper bound the summation in (\ref{eq:propeqn}) for an arbitrary splitting of the summation indices.\\
   \begin{align}
  \sum_{r=M}^{\floor{n/2}} {n \choose r}\pr[\sdcon]\nonumber&=  \sum_{r=M}^{\floor{n/\log n}} {n \choose r}\pr[\sdcon]\nonumber \\
&+\sum_{r=\floor{n/\log n}}^{\floor{n/2}} {n \choose r}\pr[\sdcon].\label{eq:gcsplit} \end{align}{}
 
 We first upper bound each term in the summation with indices in the range $M  \leq r \leq \floor{n/\log n}$.\\

\noindent \textbf{ Range 1:  $M  \leq r \leq \floor{n/\log n}$}\\ 
 \begin{align}
  &{n \choose r}\pr[\sdcon]  \nonumber \\
 &\leq \left( \mu+(1-\mu) \left(1-\dfrac{r}{n}\right)^{\K-1} \right)^{n} \\
 & =  \left( 1- (1-\mu) \left(1-\left(1-\dfrac{r}{n}\right)^{\K-1}\right) \right)^{n}
  \end{align}
For $r  \leq \floor{n/\log n}$, we have $\frac{r}{n}=\oo(1)$. Using Fact (\ref{eq:kcon_junfact}) with $x=\frac{r}{n}$ we get
   \begin{align}
  &{n \choose r}\pr[\sdcon]  \nonumber \\
&\leq \left( 1- (1-\mu)\left(1-\left(1-\dfrac{r(\K-1)}{n}+\dfrac{r^2 (\K-1)^2}{2n^2} \right)\right) \right)^{n} \nonumber \\
&= \left( 1- (1-\mu)\left(\dfrac{r(\K-1)}{n}-\dfrac{r^2 (\K-1)^2}{2n^2} \right) \right)^{n}  \nonumber \\
&= \left( 1- (1-\mu)\dfrac{r(\K-1)}{n}\left(1-\dfrac{r (\K-1)}{2n} \right) \right)^{n}  \nonumber
\end{align}
Using $r \leq n / \log n$, (\ref{eq:kcon_1pmx}) and that $\K$ is bounded above we obtain,

\begin{align}
  &{n \choose r}\pr[\sdcon]  \nonumber \\
  &\leq \left( 1- (1-\mu)\dfrac{r(\K-1)}{n}\left(1-\dfrac{ (\K-1)}{2 \log n} \right) \right)^{n}   \nonumber \\
    & \leq \exp \left\{-(1-\mu){r(\K-1)}\left(1-\dfrac{ (\K-1)}{2 \log n} \right)\right\}   \label{eq:gc_range1a} \\ 
        & = \exp \left\{-r(1-\mu){(\K-1)}\left(1- \oo(1) \right)\right\}  \label{eq:gc_range1b} \\ 
    & = \exp \left\{-{r(\kk-1)}\left(1- \oo(1) \right)\right\}.    \label{eq:gc_range1}
\end{align}
Next, we upper bound the second term in the summation (\ref{eq:gcsplit}) with indices in the range $\floor{n/\log n}+1 \leq r \leq {\floor{n/2}}$.
\\\noindent \textbf{ Range 2: $\floor{n/\log n}+1 \leq r \leq {\floor{n/2}}$}\\
Observe that
 \begin{align}
  {n \choose r}\pr[\sdcon]& \leq
 \left( \mu+(1-\mu) \left(1-\dfrac{r}{n}\right)^{\K-1} \right)^{n}\nonumber \\
 & \leq
 \left( \mu+(1-\mu) \left(1-\dfrac{r}{n}\right) \right)^{n} \label{eq:gc_range2b} \\
 & =
 \left( 1- \dfrac{r}{n} (1-\mu)\right)^{n} \nonumber \\
 & \leq \exp \left(-r(1-\mu) \right) \label{eq:gc_range2a} \\
 & =\oo(1), \label{eq:gc_range2}
  \end{align}
where (\ref{eq:gc_range2a}) follows from noting that $\K \geq 2$ and (\ref{eq:gc_range2b}) is a consequence of (\ref{eq:kcon_1pmx}). Finally, we use (\ref{eq:gc_range1}) and (\ref{eq:gc_range2}) in (\ref{eq:gcsplit}) as follows.
\begin{align}
 & \sum_{r=M}^{\floor{n/2}} {n \choose r}\pr[\sdcon]\nonumber \\
 &=  \sum_{r=M}^{\floor{n/\log n}} {n \choose r}\pr[\sdcon] \nonumber \\
 &+ \sum_{r=\floor{n/\log n}}^{\floor{n/2}} {n \choose r}\pr[\sdcon] \nonumber\\
  &\leq  \sum_{r=M}^{\floor{n/\log n}}  \exp \left\{-{r(\kk-1)}\left(1- \oo(1) \right)\right\} + \sum_{r=\floor{n/\log n}}^{\floor{n/2}} \oo(1)\nonumber\\
   &= \left(  \sum_{r=M}^{\floor{n/\log n}}  \exp \left\{-{r(\kk-1)}\left(1- \oo(1) \right)\right\}\right) + \oo(1)  \nonumber \\
      &\leq \left(  \sum_{r=M}^{\infty}  \exp \left\{-{r(\kk-1)}\left(1- \oo(1) \right)\right\}\right) + \oo(1).  \nonumber 
\end{align}
Observe that the above geometric series has each term strictly less than one, and thus it is summable. This gives
\begin{align}
      &{n \choose r}\pr[\sdcon] \nonumber \\
      &\leq \frac{\exp\{-M\left(\kk-1\right)(1-\oo(1))\}}{1-{\exp\{-\left(\kk-1\right)(1-\oo(1))\}}} + \oo(1) .
\end{align}
\myendpf

\subsection{\bf Mean node degree in $\hh$}
Let $\kk$ denote the mean number of edges that each node chooses to draw. Conditioning on the class of node~$i$, we get
\begin{align}
\kk&=\mu+(1-\mu)K_n. 
\end{align}
The probability that node~$i$ picks node~$j$  where $i,j\in \nodes$ depends on the type of node~$i$ and is given by
\begin{align}
\pr[j \in \Gamma_{n,i}]=\mu\dfrac{1}{n-1}+(1-\mu)\dfrac{K_n}{n-1}=\dfrac{\kk}{n-1}.
\end{align}

Recall that each node draws edges to other nodes independently of other nodes. Let $i \sim j$ denote the event that node~$i$ can securely communicate with node~$j$. For $i \sim j$ to occur, either node $i$ selects node $j$ or node $i$ selects node $j$ or both select each other. This gives
\begin{align}
\pr[i \sim j]&= 1 - (1-\pr[i \in \Gamma_{n,j}])(1-\pr[j \in \Gamma_{n,i}]),\nonumber \\
&=1-\left(1-\dfrac{\kk}{n-1}\right)^2,\nonumber\\
&=\dfrac{2\kk}{n-1}-\left(\dfrac{\kk}{n-1}\right)^2.
\end{align}
Consequently, the mean degree of node~$i$ can be computed as follows.
\begin{align}
\E\left[\sum_{j\in \nodes_{-i}}\nonumber \ii\{i \sim j\}\right] &= (n-1)\pr[i \sim j],\\
&=2\kk-\dfrac{\kk^2}{n-1}. \label{eq:avgnd}
\end{align}

\subsection{\bf Inhomogeneous random K-out graph with $r$ classes}

 Here, each node belongs to type-$i$ with probability $\mu_i$ for $i=1,\ldots,r$ and $\sum_{i=1}^r \mu_i=1$. Each type-$i$ nodes gets paired with $K_{i,n}$ other nodes, chosen uniformly at random from among all other nodes where $1 \leq K_{1,n} < K_{2,n} < \ldots < K_{r,n}$. Let $\pmb{K_n}$ denote $[K_{1,n}, K_{2,n},\ldots, K_{r,n}]$ and $\pmb{\mu}=[ \mu_1,\mu_2, \ldots, \mu_r ]$ with $\mu_i>0$.
\begin{corollary}
{\sl  Consider a scaling $\pmb{K_n}: \mathbb{N}_0 \rightarrow \mathbb{N}^r_0$ and a probability distribution $\pmb{\mu}=[ \mu_1,\mu_2, \ldots, \mu_r ]$ with $\mu_i>0, \ i=1,2,\dots,n$.  
and $1 \leq K_{1,n}<K_{2,n}\dots < K_{r,n}$. If $K_{r,n} \geq 2 \ \forall n$ then for the inhomogeneous random K-out graph $\mathbb{H}\left(n; \pmb{\mu}, \pmb{K}_n\right)$ with $r$ node types, we have
\begin{align}
|C_{\rm max}(n;\pmb{\mu}, \pmb{K}_n )|
  = n-\OO(1) \ \ {\rm whp.} 
  \label{eq:giant_comp_result}
\end{align}
 }\label{cor:gc_r_classes}
\end{corollary}{}

\textbf{Proof of Corollary~\ref{cor:gc_r_classes}}

The proof involves showing the existence of a coupling between the graphs $\hh$ and $\HH$ such that the edge set of $\hh$ is contained in the edge set of $\HH$. For any monotone-increasing property $\mathcal{P}$, i.e., a property which holds upon addition of edges to the graph (see \cite[p.~13]{coupling2017}) we have
 \begin{align}
    &\pr[ \HH \textrm{ has property } \mathcal{P} ]  \nonumber\\
    &\geq \pr[\hh \textrm{ has property } \mathcal{P}] \label{eq:gcmonotoneproperty}
\end{align}
It is plain that the property $ |\cmax| \geq n - M$ is monotone increasing upon edge addition. Therefore, if there exists a {\em coupling} under which 
$\hh$ is a spanning subgraph of $\HH$; i.e., if we can generate an instantiation of  $\HH$ by  adding edges to an instantiation of $\hh$, then we can use (\ref{eq:gcmonotoneproperty}) to establish this Corollary. 
Let $\tilde{\mu}$ denote $\sum_{i=1}^{r-1}\mu_i$. Consider an instantiation of an inhomogeneous random graph $\hhgc$ with two classes such that each of the $n$ nodes is independently assigned as type-1 (resp., type-2) with probability $\tilde{\mu}$ (resp., $1-\tilde{\mu}$) and then type-1 (resp., type-2) nodes draw edges to $1$ (resp. $K_{r,n}$) nodes chosen uniformly at random. From this instantiation, we can generate an instantiation of $\HH$ as follows. First, let each type-1 node be independently reassigned as type-$i$ with probability $\frac{\mu_i}{\tilde{\mu}}$ for $i=1,2,\dots,r-1$. Next, for $i=2,\dots,r-1$, let each type-i node pick $K_{i,n}-1$ additional neighbors that were not chosen by it initially. After these additional choices are made, we draw an undirected edge between each pair of nodes where at least one picked the other. Clearly, this process creates a graph whose edge set is a superset of the edge set of the  realization of $\hhgc$ that we started with. In addition, in the new graph, the probability of a node picking $K_{i,n}$ other nodes (i.e., being type-$i$) is given by  $\tilde{\mu} \frac{\mu_i}{\tilde{\mu}} = \mu_i$, for $i=1,2,\dots,n$. We thus conclude that the new graph obtained constitutes a realization of $\HH$. Since, the initial realization of  $\hhgc$ was arbitrary, this establishes the desired coupling argument and we conclude 
  that (\ref{eq:gcmonotoneproperty}) holds for the property $ |\cmax| \geq n - M$. The proof of Corollary \ref{cor:gc_r_classes} is now complete.

  \myendpf


\bibliographystyle{IEEEtran}
\bibliography{IEEEabrv,references}
\end{document}